%% file: main.tex
\documentclass[journal,comsoc]{IEEEtran}

\usepackage{url} 
\usepackage{balance}
\usepackage{float}
\usepackage{amsmath,amssymb,ctable}  

\usepackage{subcaption}
\usepackage{graphicx}

\usepackage{algorithm}
\usepackage{algpseudocode}

\begin{document}

	\title{Intelligent Routing Algorithm over SDN: Reusable Reinforcement Learning Approach}
	
	\author{Wang Wumian, Sajal Saha, Anwar Haque, and Greg Sidebottom

		\thanks{W. Wumian, and A. Haque are with the Department of  Computer Science, Western University, London, ON N6A 5B7 Canada e-mail: \{wwang457, ahaque32\}@uwo.ca.}
		\thanks{S. Saha is with the Department of Computer Science, University of Northern British Columbia, Prince George, BC V2N 4Z9 Canada e-mail: \{sajal.saha\}@unbca.ca}
		\thanks{G. Sidebottom is with the Juniper Networks, Kanata, ON K2K 3E7 Canada email: \{gsidebot\}@@juniper.net}
	}
	
	
	\maketitle
	\input{abstract}
	
	\section{Introduction}
    \input{introduction}
    
    \section{Problem Statement and Related Work}
    \label{sec:related-work}
    \input{related-work}
    
    \section{Proposed Methodology}
    \label{sec:methodology}
    \input{methodology}

    \section{Experiment Setup and Result Analysis}
    \label{sec:experiments}
    \input{experiments}

    \section{Conclusion}
    \label{sec:conclusion}
    \input{conclusion}

	\balance
	\bibliographystyle{ieeetr} 
	\bibliography{ensemble}
	
	\begin{IEEEbiography}
    [{\includegraphics[width=1in,height=1.25in,clip,keepaspectratio]{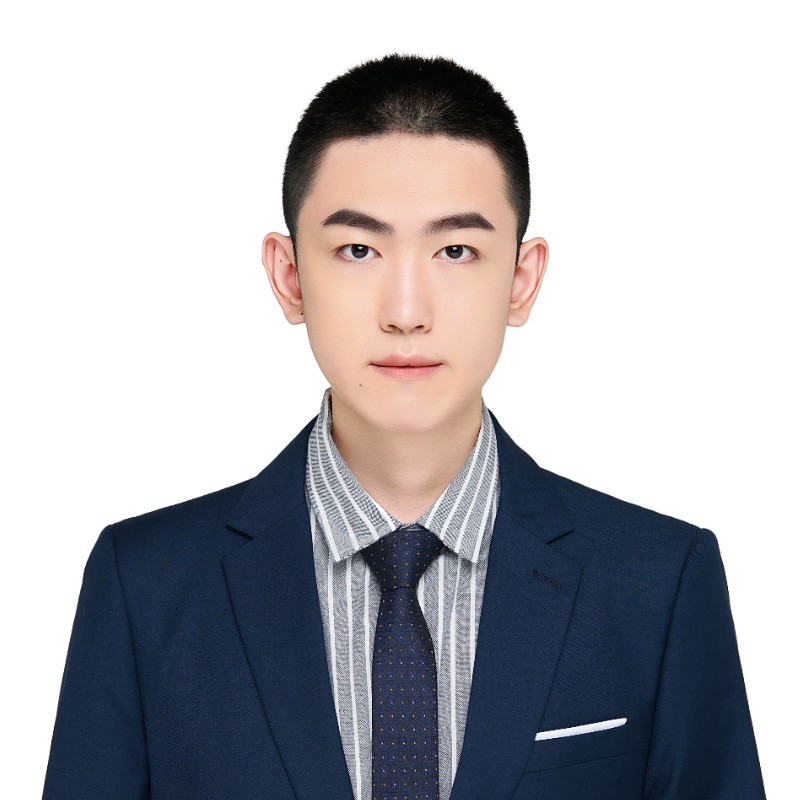}}]
  {Wumian Wang,} earned a bachelor’s degree in microbiology \& immunology from the University of British Columbia between 2014 and 2019. Following that, Wang pursued a master’s degree in computer science at The University of Western Ontario from 2020 to 2022, during which he won a research scholarship for the year 2021-2022. Alongside his studies, Wang worked as a teaching assistant in a computer science course in 2022. 
    \end{IEEEbiography} 
    
    \begin{IEEEbiography}
    [{\includegraphics[width=1in,height=1.25in,clip,keepaspectratio]{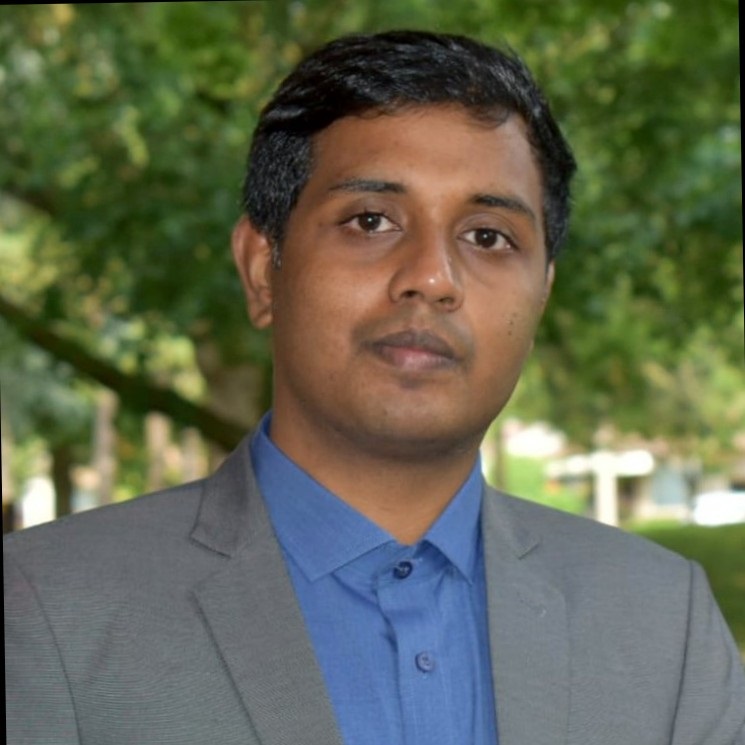}}] {Sajal Saha, Ph.D.} is an Assistant Professor in the Department of Computer Science at the University of Northern British Columbia, Canada. His research interests are primarily focused on Internet Traffic Analysis, Cyber-attack Detection, and Data Mining, with special emphasis on traffic prediction, anomaly detection, and association rule mining. He completed his Ph.D. at Western University, Canada, where his research contributed significantly to efficient traffic forecasting and cyber-attack detection. Alongside his academic contributions, he has also gained considerable industry experience as a Research Fellow at Juniper Network and Software Engineer at Samsung Electronics.
    \end{IEEEbiography}
    
     \begin{IEEEbiography}
    [{\includegraphics[width=1in,height=1.25in,clip,keepaspectratio]{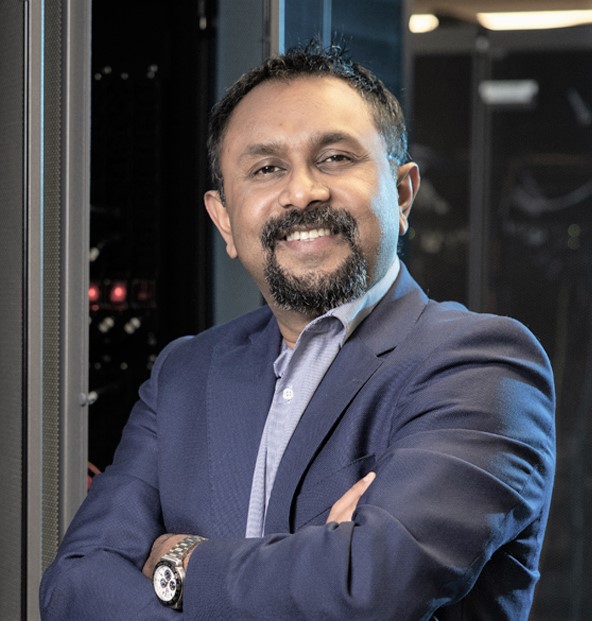}}] {Dr. Anwar Haque} is an Associate Professor and the Undergraduate Chair of the Department of Computer Science at Western University, Canada. Dr. Haque holds a Ph.D. in Electrical \& Computer Engineering and a Master’s degree in Computer Science, both from the University of Waterloo. Before joining Western, he was an Associate Director at Bell Canada. Dr. Haque’s primary research areas of interest include next-gen communication networks, IoT, cyber-security, and applied AI focusing on developing secure and resilient autonomous cyber-physical systems (CPS). He has authored/co-authored more than 100 peer-reviewed research publications in leading journals and conferences, holds several patents/licenses, and supervised more than 100 HQPs. His collaborative research grants are valued at more than \$15 million. At Western, he served as the industry expert-in-residence in the Faculty of Science, a member of Western’s Senate, and on the inaugural advisory committee for the newly established Bell-Western 5G Research Centre. Additionally, Dr. Haque is the founder and CEO of Bamboo Innovations Inc., a successful tech venture in the smart systems industry.
    \end{IEEEbiography}
    
    \begin{IEEEbiography}
    [{\includegraphics[width=1in,height=1.25in,clip,keepaspectratio]{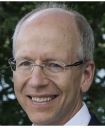}}] {Greg Sidebottom} received his PhD. degree in Computer
Science from Simon Fraser University. He is a Principal Engineer at Juniper Networks. He is the architect of Juniper’s industry pioneering Software Defined Network and Network Function Virtualization products. Greg is currently advancing the state of the art in segment routing and peer engineering with Juniper’s Northstar WAN SDN controller.
    \end{IEEEbiography}

\end{document}

%% file: abstract.tex
\begin{abstract}
Traffic routing is vital for the proper functioning of the Internet. As users and network traffic increase, researchers try to develop adaptive and intelligent routing algorithms that can fulfill various QoS requirements. Reinforcement Learning (RL) based routing algorithms have shown better performance than traditional approaches. We developed a QoS-aware, reusable RL routing algorithm, RLSR-Routing over SDN. During the learning process, our algorithm ensures loop-free path exploration. While finding the path for one traffic demand (a source destination pair with certain amount of traffic), RLSR-Routing learns the overall network QoS status, which can be used to speed up algorithm convergence when finding the path for other traffic demands. By adapting Segment Routing, our algorithm can achieve flow-based, source packet routing, and reduce communications required between SDN controller and network plane. Our algorithm shows better performance in terms of load balancing than the traditional approaches. It also has faster convergence than the non-reusable RL approach when finding paths for multiple traffic demands. \\

\textbf{Keywords: Reinforcement Learning (RL), Traffic Routing, Software Defined Networking (SDN), Segment Routing (SR), Quality of Service (QoS).}

\end{abstract}

%% file: introduction.tex
Over the past decade, the number of people who have access to the Internet has been steadily increasing. CISCO’s report predicts that by 2023, there will be 5.3 billion Internet users, which is about two third of the World’s population \cite{1}. Routing, which is the process of determining and forwarding packets from their source to destination, is vital for the transmission of data and information between network of networks, a.k.a., the Internet \cite{2}. Traditionally, traffic routing relies on protocols like OSPF, which aims to find the shortest path between source and destination \cite{3}, or Routing Information Protocol (RIP), which is a distance-vector based algorithm that mainly considers hop count \cite{4}. OSPF is one of the most widely used Interior Gateway Protocol, and the shortest paths between source and destination pairs are calculated based on assigned weights to every link of a network \cite{5}. To obtain good network performance instead of only considering number of hops along a routing path, the links weights should be optimized. However, a previous study has proved that optimizing OSPF weights is a NP-complete problem \cite{6}.

To provide better network performance for increasing demands over a variety of applications, new network architectures and traffic routing methods have been proposed. Software Defined Networking (SDN) is an architecture that decouples the control plane and data plane of a router \cite{7, 8, 9}. Traditional network routers have their own control and data plane \cite{8}, and in a protocol like OSPF, each router maintains its own forwarding table \cite{3}. For SDN, the control plane is in a logically central controller which has a global view of the network; and the SDN controller enables better interaction with applications, Quality of Service (QoS) provisioning, network monitoring and so on \cite{7, 8}. SDN can also be applied to the wireless network in addition to traditional Wide Area Network \cite{9}. Segment Routing (SR) \cite{10} is a source packet routing architecture in that each packet encodes a list of Segments in its header to guide the packet travel through the specified path \cite{10, 11}. As stated in \cite{12}, SR’s ability to control packet forwarding can provide better network programmability, fast reroute, load balancing etc. In addition to the new routing architecture, researchers also try to apply Reinforcement Learning (RL) in traffic routing \cite{13}. One recent implementation of RL in SDN can also be found in \cite{14}.

\subsection{Motivation and Objectives}
Reinforcement Learning has been studied to develop adaptive, intelligent routing algorithms over the past three decades \cite{13}. It has shown extraordinary flexibility to combine with other techniques for traffic routing over a variety of network types, with different QoS requirements \cite{13, 15}. Compared with traditional routing methods which are based on assumptions about network conditions, applying RL allows routing algorithms to automatically learn the dynamic of the network, such as traffic flows, link quality and so on \cite{15}. Therefore, based on RL’s self-learning ability, some of the previous research’s objectives are to improve service quality to end users, while optimizing network resources \cite{15}. However, many previously proposed algorithms do not fully exploit the global view of SDN controller, or the ability to control the whole forwarding path (i.e., flow-based routing) by combining with technique like SR \cite{15}. In addition, many previous works only focus on one traffic demand at a time, and some require pre-defined algorithms (like OSPF) or initial paths as input \cite{15}. 

Our objective is to develop a RL-based routing algorithm for SDN, such that it can address previous work’s limitations. Our algorithm should be light-weighted, efficient, and allow users to customize their QoS requirements for definition of users’ preferred routing paths. While the algorithm aiming to minimize costs during the path finding process, the final routing path from the algorithm should be as good as possible in terms of satisfying user defined QoS requirements. 

\subsection{Contribution}
We developed a RL-based, SR-based, QoS-aware routing algorithm for SDN, RLSR-Routing. The algorithm considers various QoS factors, which can be customized by users, to find a user preferred path for traffic demands. The main contributions of this research are listed below:
\begin{itemize}

    \item We modified State-Action-Reward-State-Action (SARSA), an on-policy RL method \cite{13}, so that our routing algorithm aggregates action selections first. Such modification can further reduce the number of message exchanges (e.g., sending packets, receiving QoS info) required between SDN controller and network switches. 

    \item In addition, during the learning process, our algorithm ensures that no packet will be stuck in a loop, which means that packets either reach the destination or stop being forwarded if all next-hop nodes have been visited by themselves. In other words, a packet will be sent to any node in the network at most once. 

    \item We divided an action’s reward into local and global rewards. Local rewards are used for finding a path for one traffic demand, whereas global rewards are purely based on network status like link utilizations. Global rewards can be used to initialize local Q-table to speed up algorithm convergence.

    \item RLSR-Routing does not require prior knowledge of the network. In addition, it does not require pre-defined paths or initial link weights as a starting point. The algorithm can self-learn network status and assign a list of traffic demands over the network in sequence. 

\end{itemize}

We compared our RLSR-Routing with a routing algorithm currently used by one of the major telecom solutions providers. A mesh topology network is setup with randomly generated traffic to be placed over the network. We compared the quality of selected paths for generated traffic, in terms of maximum link utilization. Our experiment result shows that the proposed RLSR-Routing approach contributes to minimizing link utilization in the network. In addition, when tested on a sequence of traffic demands, the proposed RLSR-Routing approach speeds up the algorithm convergence by applying knowledge of the network from its previous learning processes.

The structure of this paper is outlined below. In section II, we define our problem, network environment, and introduce previous researchers that use RL approach to solve traffic routing problems. Section III presents our methodology, including overall architecture, pseudocode of RLSR-Routing, and rationale behind our algorithm designs. In section IV, we summarize our experiment results, including a comparative study with a non-RL routing algorithm. In section V, we conclude our study and propose potential updates and objectives for future work.

%% file: related-work.tex
In this section, we formally define our research problem and present a literature review of recent work related to our problem and existing solutions. This chapter also highlights the current research gap and sets transitions to introducing our proposed methodology.

\subsection{Problem Statement}
We focused on traffic routing in Software Defined Networking (SDN); therefore, we define our network as a set of SDN switches with an SDN controller that has a global view of the network. This allows the controller to know the topology of the network, including every node and link. The controller has dedicated links to communicate with network nodes, but neither the controller nor these links are used for routing packets between the network’s end users. Our routing algorithm interacts with the SDN controller to perform path calculation and installs defined routing paths into the network.

From the routing algorithm’s point of view, a network can be abstracted as a directed graph, \(G(V, E)\), where \(V\) is the set of nodes that each corresponds to a physical device (SDN switch), and \(E\) is the set of links that connects nodes. \(l_{i,j}\) denotes the unidirectional link from node \(i\) to \(j\) where \(i, j \in V\). We define a traffic demand as a flow of data transmitted from a source node to a destination node. One source and destination pair can have multiple traffic demands, each with a specified amount of data as demand traffic (expressed in bits/second). However, each traffic demand can only be assigned one routing path, which is defined as a list of nodes that connects from source to destination.

Each link and node of a network has a maximum capacity. For a link, we use the term “maximum bandwidth” (in bits/second) to represent the maximum traffic that can travel through the link, and “used bandwidth” (in bits/second) to represent the current amount of data on the link. A node’s processing rate (in bits/second) is the maximum amount of incoming data it can handle. In addition, all links have a field called “link reliability”, which varies from 0\% to 100\%. The higher the link reliability, the lower the chances of link failure will occur.

Based on the above settings, our research objective is to give a network with SDN architecture with all nodes supporting Segment Routing (SR) and develop an RL-based routing algorithm that can find user-preferred paths for traffic demands. Our definition of ‘user-preferred’ paths are those paths found by our routing algorithm based on weights of one or more Quality of Service (QoS) parameters, which are adjusted by our algorithm’s users. The following assumptions are made:

\begin{itemize}
    \item Communications between SDN controllers and network switches are through dedicated links, in a reliable transfer manner.
    \item SDN controllers are not involved in routing packets that originated from the network’s end users. For better visibility, sometimes we do not draw the controllers and those dedicated links on network topology diagrams in this research.
    \item All SDN switches of the networks considered in this research support SR, i.e., they can handle packet’s SR header and execute segments properly.
    \item All nodes are reliable, and if total incoming traffic does not exceed a node’s processing rate; no packets will be dropped by nodes.
    \item For each pair of adjacent nodes \(i-j\) that connects with each other, at most two links exist: link \(i \rightarrow j\) and link \(j \rightarrow i\), that directly connect \(i\) and \(j\).
    \item If a link’s used bandwidth does not exceed its maximum bandwidth, and link failure does not happen, no packet will be lost when traveling on that link.
\end{itemize}

\subsection{Related Work}

\subsubsection{Reinforcement Learning on Non-Software Defined Networks}
Similar to \cite{24}, several recent research focused on RL approach for routing in WSNs \cite{28, 29, 30}. One character of WSNs is network devices are often powered by battery, therefore, energy consumption or energy conservation is an important factor to be optimized when designing RL-based routing algorithm. Both \cite{28, 29}’s work involves exchanging information related to energy conservation, which is accomplished by adding extra fields in data packets. In \cite{28}, the sender node’s current Q-value is sent to a packet’s next hop node, along with information about previous nodes energy level. In \cite{30}, the authors combine Q-learning and transmission gradient for optimizing energy consumption over the WSNs.  For transmission gradient, each node maintains an estimated number of transmissions for sending a packet from the node to the base station \cite{30}. Estimating the number of transmissions and energy level are used for calculating rewards; and nodes use status packets to send their Q-value and estimated number of transmissions to other nodes \cite{30}. 

Another type of network which many studies focused on is Ad-hoc Networks (ANETs). In general, ANETs do not have a centralized controller, nodes are wirelessly connected, and network topology is dynamic. Mobile Ad-hoc Networks (ANETs) is a subtype of ANETs that nodes have mobility; common types of MANETs include VANETs for vehicles moving in certain area like a city \cite{15}. In this review, all 6 recent papers which focus on ANETs used packet loss/retransmission rate/delivery ratio as one of metric for performance evaluation \cite{31, 32, 33, 34, 35, 36}. Unlike early work that is only based on RL, many of these ANETs related papers combine RL with other techniques or have additional steps in addition to RL algorithm. Authors in \cite{31} proposed QLMAODV, a combination of Q-learning and AODV protocol. AODV is defined in RFC3561, it is a routing protocol for MANETs, and one notable feature is that AODV ensures loop free routing \cite{37}. QLMAODV considers stability when Q-learning evaluates a path to achieve a stable route \cite{31}. In \cite{32}, the authors proposed PFQ-AODV, an extension of their previous work in \cite{38}. PFQ-AODV also combines AODV with Q-learning. In addition, it considers bandwidth, node’s movement, and link quality in route selection by fuzzy logic \cite{32}. Compared with their previous work, PFQ-AODV can use neighbor information, in addition to position information, to calculate vehicle movements. QGeo proposed in \cite{33} also considers nodes mobility: each node exchanges their location information periodically through HEELO packets \cite{33}. 

Authors in \cite{34} proposed a routing algorithm based on fuzzy logic, game theory and RL. Fuzzy logic groups vehicles (as nodes) into clusters and selects cluster heads, and game theory coordinates commutations between nodes and their cluster heads. Nodes are enforced to use cluster heads for multi-hop transmission and RL is used for path evaluation. Similarly, Q2-R proposed in \cite{35} has additional bootstrapping steps before RL routing. Q-learning starts with initial paths that discovered from bootstrapping steps, and agent gradually improves paths’ Q-value during learning process. A hierarchical routing scheme is proposed in \cite{36}, which first divides geological area into grids without the selection of cluster heads. Q-learning is used for selecting next grid and can be used to select the next hop (vehicle) inside the optimal grid.

Author in \cite{39, 40} focused on Wireless Mesh Networks (WMNs), in which nodes with multiple interfaces serve as a gateway that connects other nodes to the Internet \cite{15, 40}. In \cite{39}’s framework, each node predefined a set of routing algorithms, and Q-learning is used for adaptively selecting routing algorithms to execute. In \cite{40}, gateway nodes periodically broadcast their traffic load, each node first selects the gateway router, then uses RL to select next-hop node to send packets to the selected gateway. Authors in \cite{41} focused on multi-hop wireless networks, particularly for video streaming. The proposed RLOR algorithm combines RL and Opportunistic Routing, which exploit the broadcast nature of the wireless network and determines packets’ next-hop on the fly.

\subsubsection{Reinforcement Learning on Software Defined Networks}
As illustrated in \cite{14}, when acting as an RL agent, SDN controller(s) can find optimal path from source to destination, in addition to next-hop node. To address the scalability issue of a single controller, the authors of \cite{42} proposed QAR, a QoS-aware routing algorithm on hierarchical SDN. Nodes are divided into subnets, and each subnet has a domain controller, assisted by one or more “slave” controllers. When a routing request’s source and destination are at different subnets, the main controller with a global view calculates a subnet path that connects the source and destination’s subnets. Then each subnet along the path runs RL on their domain controller to travel packets within subnets, and such path finding can be done in parallel. Another notable feature of QAR is that it uses SARSA, an on-policy RL algorithm that compared with Q-learning, updates of a state-action pair’s Q-value depends on action performed on the next state. \cite{43, 44} is also based on SARSA, but on a non-hierarchical SDN. VS-routing in \cite{43} applies a variable \(\varepsilon\)-greedy for action selection, such that the probability of randomly selecting an action instead of a greedy approach varies with hop count and a dynamic factor \cite{43}. In addition, VS-routing has no immediate rewards, Q-value updates are based on the next state’s associated Q-values and \(\varepsilon\). A multiple controllers SDN is considered in \cite{44}, and its objective is load-balancing for communication between nodes and controllers. also improved \(\varepsilon\)-greedy action selection, by using the Bayesian method: during exploration, agent tends to select frequently selected actions.

For Q-learning based routing algorithms in SDN, we also observed diversity in terms of network architecture and algorithms’ workflows. Authors of \cite{45} focused on congestion control in SDN. Learning is not based on selected actions and immediate rewards but update all edges reward in each episode of the routing algorithm. Congestion identification is based on the current bandwidth of potential bandwidth, same for updating each edge’s corresponding reward. In \cite{46}, Q-learning based routing is used for load-balancing in Wireless SDN. In general, a Wireless SDN still has a central controller and set of SDN switches, in addition, SDN base stations are used to connect with SDN mobile devices. States are assigned to users and depend on whether users’ demands are satisfied, actions are connecting every user’s traffic flow to a base station, and rewards are calculated based on available resource and a fairness function. QR-SDN in \cite{47} also adapts flow-based routing. A flow is defined as data transmission between a given source and destination pair, and each flow is assigned one path for routing traffic. QR-SDN will not compute the path for a given flow, instead, it selects a path from a set of pre-calculated possible paths. During the learning process, QR-SDN can take a list of flows, change path assignment for one or more flows in each episode, and gradually learns optimal path assignments.

Author in \cite{48, 49} use Q-learning, but on a “distributed” SDN. In \cite{48}, the authors implemented a “distributed” SDN over several computers (PC), with each PC containing a part of SDN network with one controller. Links are initially assigned weights, and the objective of Q-learning is to optimize link weights assignments. SDCIV proposed in \cite{49} aims to achieve adaptive routing in SDN based Internet of Vehicles (IoV). The logical central controller is made up with a SDN controller cluster, and thus we classify it as a “distributed” operation mode. SDCIV defines states as average vehicle speeds and densities, and the learning agent gradually learns the best routing protocol to execute for different states. Similar to \cite{39}, RL is not used for routing or path finding directly, but to find an appropriate protocol to execute from pre-defined routing algorithms.

\subsubsection{Deep Reinforcement Learning based Routing}
Using RL in traffic routing started in 1990s, in contrast, DRL based approach is a relatively emerging field. Fig. 3 shows a number of published papers related to DRL and routing between 2017 and 2021, from Web of Science database. Since this literature review focuses on traffic routing, we do not cover how neural network is trained in different DRL based routing algorithms.

In \cite{50}, the authors proposed DROM, a routing optimization algorithm for SDN. DROM takes a traffic matrix of current network load as input, during the learning process, the DRL agent changes some links’ weight so that some traffics alters their routing path \cite{50}. WA-SRTE in \cite{51} works on a partially deployed Segment Routing IPv6 network, and it has two phases. In offline network design phases, WA-SRTE uses DRL to optimize OSPF link weights and SR-enabling nodes’ deployment based on historical traffic data. Once the offline phase is complete, link weights and SR nodes deployment is fixed. In the online routing optimization phase, WA-SRTE uses other techniques like linear programming to optimize routing paths. ENERO proposed in \cite{52} also combines Segment Routing and its objective is also to minimize maximum link utilization. Taking initial OSPF weights and a set of traffic demands as input, ENERO uses DRL to assign each traffic demand to an intermediate SR node to optimize links utilizations. To further improve the result, local search (LS) is applied to DRL’s output in ENERO \cite{52}.

In \cite{53}, the authors applied Deep Q-learning for routing in data center SDN. The authors classified different traffic flows into two types: the mice-flows and the elephant-flows based on the amount of data and duration of traffic. \cite{53} thus builds two Deep Q-networks to assign paths for two traffic flows, respectively, and each flow type has a different QoS objective. \cite{54} proposed a hierarchical deep double Q-routing algorithm, which groups nodes into different clusters at different hierarchal levels, each cluster with a group leader. During the recursive route-finding process, the source’s routing request is sent to the highest level’s cluster leader, such that the source and destination nodes are at two different sub-clusters. The cluster leader uses DRL to select one link that connects two sub-clusters, and the process is repeated until a whole path is built. The double Q-learning technique is adapted from \cite{55}; when combined with deep learning, two Deep Q-networks are interchangeably used for action selection and evolution.

\subsection{Analysis and Current Research Gap}
Most of our reviewed papers proposed an RL-based routing algorithm, although the DRL approach has attracted researchers’ interest dramatically in the past 5 years. Among studies that focus on non-SDN networks, 85.7\% are on kind of wireless network and/or Ad-hoc network. In addition to latency-related factors (like delay or hop-count), RL-based approaches can consider a variety of factors that affect network performance. However, when working on a non-SDN network like WSNs or ANETs, applying RL-based routing may generate additional commutations overhead. For example, \cite{28,29,30} routing algorithms require exchanging energy consumption or device residual energy information among network nodes. With SDN controllers’ global view of their networks, or a hierarchical network with some nodes being cluster heads (like \cite{34, 54}) could reduce commutations overhead placed by RL-based routing over a network.

61.5\% of reviewed papers proposed a composite routing protocol, where RL or DRL is part of the whole routing protocol. Common reasons of combining RL with other techniques are: 1. To speed up the algorithm’s convergence (like QAR in \cite{42}); 2. To provide additional control, such as loop free routing \cite{31, 32} or source packet routing \cite{52, 51}.  It should be noticed that for some studies on SDN, like QAR, their routing algorithm does not guarantee loop-free routing during the learning phase, even though SDN controller(s) are RL agents \cite{42}. The usage of RL is not limited to finding the best next-hop node to forward a packet. RL agents can learn the whole path from a source to a destination or assign a path to traffic flows from a set of possible paths. In addition, several studies focus on OSPF optimization by using RL approach, which we define as the indirect usage of RL in traffic routing. Most papers’ framework only focuses on forwarding one-packet, or one source-destination pair’s traffic. None of our reviewed literature, which is based on RL over SDN, use RL to directly find the path for a set of traffic flows in parallel.

In summary, RL has been studied to develop adaptive, intelligent routing algorithms over the past three decades. It has shown extradentary flexibility to combine with other techniques for traffic routing over a variety of network types, with different QoS requirements. However, currently proposed algorithms do not fully exploit the global view of the SDN controller, or the ability to control the whole forwarding path (i.e., flow-based routing) by combining with a technique like SR. Additional work needs to be done in order to develop a RL-based, QoS aware flow-based routing over SDN, such that RL agent can find path for multiple traffic demands in parallel. Alternatively, while RL agent still focuses on one traffic demand at a time, the agent can reuse the network status knowledge it has learned previously to speed up algorithm convergence.

%% file: methodology.tex
As the number of Internet users and types of services that rely on networking increases, traditional distance vector or shortest path based algorithms may not be able to provide optimal network performance. Previous studies demonstrate the potential of using RL or DRL to develop an intelligent routing algorithm that is adaptive to dynamic network topology and fulfills a variety of QoS requirements. To the best of our knowledge, existing approaches do not fully exploit SDN architecture combined with a flow-based source packet routing paradigm. In this research, we propose RLSR-Routing, an RL-based routing algorithm over SR-enabled SDN architecture to address the current research gap. RLSR-Routing reduces network operation costs during the path-finding process and can have faster convergence than the previous RL-based approach. This section described the RLSR-Routing framework and the rationale behind RLSR-Routing’s design. 

\subsection{RLSR-Routing Architecture}
\label{subsec:rlsr-routing architecture}

\begin{figure}[!htbp]
	\centering 
	\includegraphics[width=0.4\textwidth]{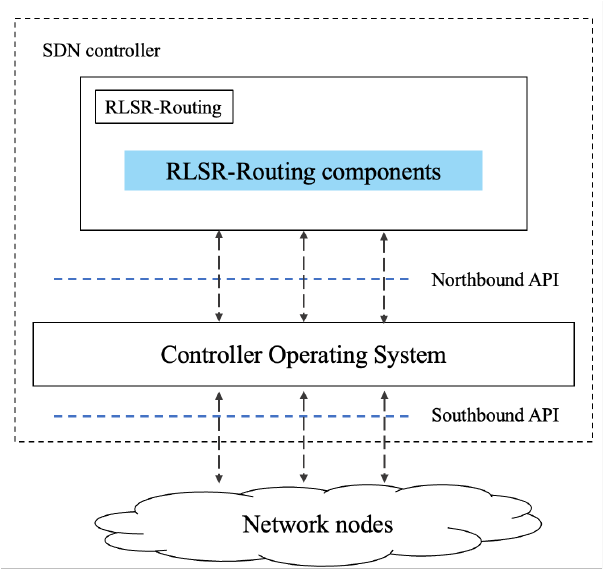}	
    \caption{High-level Overview of RLSR-Routing Architecture} 
	\label{fig:rsrl_arachitecture}
\end{figure}

\begin{figure}[!htbp]
	\centering 
	\includegraphics[width=0.4\textwidth]{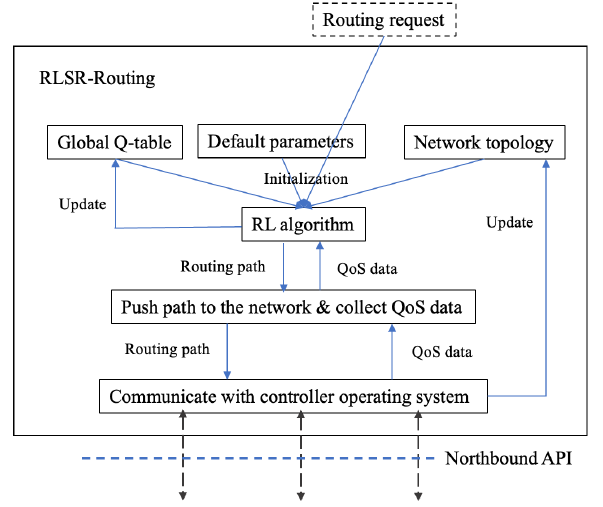}	
    \caption{High-level Overview of RLSR-Routing Components} 
    \label{fig:rsrl_component}
\end{figure}

Fig.\ref{fig:rsrl_arachitecture} presents an abstract view of our framework. RL routing algorithm is one component of the SDN controller, and it relies on the SDN control plane to provide network topology and link state information. Although different literature may categorize routing as an application plane or control plane’s function \cite{2, 7}, such difference does not affect RLSR-Routing’s framework. Because in either case RLSR-Routing does not directly interact with network devices. Fig. \ref{fig:rsrl_component} describes RLSR-Routing’s components in greater detail. These components can be classified into three categories:

\begin{enumerate}
    \item \textbf{Storage of information:} such as \textit{Global Q-table} which represents previously learned network status, \textit{Default parameters}, and \textit{Network topology}.
    \item \textbf{RL algorithm:} this is the central part of \textit{RLSR-Routing}. It finds paths for specified traffic demands, as well as learns network status during execution.
    \item \textbf{Communication:} these components interact with the SDN controller to help other components send instructions to the SDN controller, and to collect network information like QoS data, and topology information from the SDN controller.
\end{enumerate}

\begin{figure}[!htbp]
	\centering 
	\includegraphics[width=0.4\textwidth]{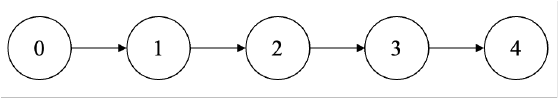}	
     \caption{Trivial Topology 1 (T1)} 
	\label{fig:trivial_topology_1}
\end{figure}

\subsection{State, Action, Reward, and Modifications}
\label{subsec:state-action-reward}

The RL algorithm used in RLSR-Routing can be viewed as an extension to previous work, particularly QAR from \cite{42}. RLSR-Routing applies modified SARSA, an on-policy RL algorithm and a complete Markov Decision Process is included in RLSR-Routing. In this subsection, we define the state space, the action space, and rewards’ QoS considerations for RL algorithm used in RLSR-Routing. We also explain the rationale for two major modifications we made in our RL algorithms.

\subsubsection{State and Action Space}
RLSR-Routing focuses on routing for one traffic demand at a time: during each episode of algorithm execution, RLSR-Routing instructs the SDN controller to send one packet from the traffic demand's source to its destination. Based on the above settings, our definition of state space and action space is as follows.

\begin{itemize}
    \item \textbf{State space:} let \(S\) denotes the set of all possible states of a network. \(|S| = |V|\) where \(V\) is set of nodes of the network. A state \(s \in S\) means node \(s\) is holding the packet. Notations \(s_t\) and \(s_{t+1}\) are also used, which represent the state at time \(t\) and time \(t+1\), respectively.
    
    \item \textbf{Action space:} let \(A\) denotes the set of all possible actions that can be performed over a network. \(|A| = |E|\) where \(E\) is set of edges of the network. An action \(a_{i,j} \in A\) represents sending the packet along link \(a_{i,j}\), from its source node \(i\) to its destination node \(j\). Notations \(a_t\) and \(a_{t+1}\) are also used, which represent selected action at time \(t\) and time \(t+1\), respectively.
    
\end{itemize}

Fig. \ref{fig:trivial_topology_1} shows a trivial topology \(T_1\) with 5 nodes, and 5 unidirectional links (followed by the arrows). The SDN controller and its connections with nodes are not shown in the figure. If a packet is sent from node 0 to node 5, at time \(t_0\), node 0 holds the packet, the chosen action is \(l_{0,1}\). At time \(t_1\), the packet reaches node 1, and therefore the state at \(t_1\) is node 1 holding the packet.

\subsubsection{QoS Considerations for Actions’ Rewards}

Previous studies have demonstrated the flexibility of RL-based routing that multiple QoS metrics can be used to calculate an action's reward. In RLSR-Routing, we considered five QoS parameters which can be categorized into three groups: latency; reliability; and load balancing. Assuming the performed action is sending one packet from node \(i\) to node \(j\) through link \(l\), the calculation of each QoS parameter's reward is presented below. For latency-related QoS metrics, we considered "number of hops" and "transmission rate".

\textbf{Number of hops related reward (\(R_{\text{hop}}\))} is calculated as follow:
\begin{equation}
    R_{\text{hop}} = \frac{1}{\text{number of hops}}
\end{equation}

\textbf{Transmission rate related reward (\(R_{\text{transmission}}\))} reflects the transmission delay of a node, and it is calculated as follows. The higher a node's processing rate, the less time it requires to transmit all bits of a packet to the link. Therefore, the related action will receive a higher reward.
\begin{equation}
    R_{\text{transmission}} = \frac{2}{\pi} \times \arctan(\text{node } i \text{ processing rate})
\end{equation}

For reliability-related QoS metric, we considered 1) link reliability. Since, in our assumption, all network switches and SDN controllers are 100\% reliable, only the possibility of link failure is considered. Calculation of link reliability-related reward (\(R_{\text{reliability}}\)) is:
\begin{equation}
    R_{\text{reliability}} = \text{link } l's \text{ reliability value}
\end{equation}

For load balance-related QoS metrics, we considered 1) traffic intensity, 2) link utilization. For these two metrics, both current and estimated (i.e., estimated traffic intensity and link utilization after placing the traffic demand’s flow on the link) values are considered. The calculation of current traffic intensity (\(R_{\text{inten}}\)) and estimated traffic intensity (\(R_{\text{inten-est}}\)) related reward are shown below:

\begin{equation}
    R_{\text{inten}} = 1 - \frac{\text{Node } j \text{ current total incoming traffic}}{\text{Node } j \text{ processing rate}}
\end{equation}
\begin{equation}
    R_{\text{inten-est}} = 1 - \frac{\text{Node } j \text{ estimated total incoming traffic}}{\text{Node } j \text{ processing rate}}
\end{equation}

Calculation of current link utilization (\(R_{\text{util}}\)) and estimated link utilization (\(R_{\text{util-est}}\)) related reward are shown below:
\begin{equation}
    R_{\text{util}} = 1 - \frac{\text{Link } l \text{ current used bandwidth}}{\text{Link } l \text{ maximum bandwidth}}
\end{equation}
\begin{equation}
    R_{\text{util-est}} = 1 - \frac{\text{Link } l \text{ estimated used bandwidth}}{\text{Link } l \text{ maximum bandwidth}}
\end{equation}

Eq. 1 – 7 ensure that all QoS-related rewards values are no greater than 1. The closer a reward’s value is towards 1, the better quality of the performed action is in terms of the reward. Take T1 as an example; suppose a packet is sent from node 0 to 4 following the path $0\to1\to2\to3\to4$. All links have $10\,\text{Mb/s}$ maximum bandwidth, $95\%$ of the time are working; all nodes have $50\,\text{Mb/s}$ processing rate. Currently, there is $5\,\text{Mb/s}$ traffic sending from node $3$ to node $4$, and the traffic demand from node $0$ to $4$ has $0.5\,\text{Mb/s}$ estimated traffic. If we evaluate the action of sending packet from node $3$ to $4$, the corresponding rewards are:

\begin{itemize}
    \item $R_{\text{hop}} = \frac{1}{4} = 0.25$ (Since link $l_{3,4}$ is the fourth hop the packet has travelled).
    \item $R_{\text{transmission}} = \frac{2}{\pi} \times \arctan(50\,\text{Mb/s}) = 0.9873$ (Round to $4$ decimals).
    \item $R_{\text{reliability}} = 95\% = 0.95$.
    \item $R_{\text{inten}} = 1- \frac{5\,\text{Mb/s}}{50\,\text{Mb/s}} = 0.9$ (Currently, there is $5\,\text{Mb/s}$ traffic towards node $4$).
    \item $R_{\text{inten-est}} = 1- \frac{5+0.5\,\text{Mb/s}}{50\,\text{Mb/s}} = 0.89$ (After placed demand's traffic, node $4$ are expected to have $5.5\,\text{Mb/s}$ total incoming traffic).
    \item $R_{\text{util}} = 1- \frac{5\,\text{Mb/s}}{10\,\text{Mb/s}} = 0.5$.
    \item $R_{\text{util-est}} = 1- \frac{5+0.5\,\text{Mb/s}}{10\,\text{Mb/s}} = 0.45$.
\end{itemize}

\subsubsection{RL Modification (Aggregate Action Selection)}
Here, we describes and explains one of the major modifications we made for SARSA used in RLSR-Routing - aggregate action selection of one episode. Based on our literature review, the two most commonly used RL algorithms in traffic routing are Q-learning and SARSA. Both Q-learning and SARSA follow a similar workflow during each episode of algorithm execution.

\begin{enumerate}
    \item At time $t$, from current state $s_t$, select an action $a_t$ based on action selection policy.
    \item Perform action $a_t$.
    \item Observe and/or calculate action's reward, and new state $s_{t+1}$ at time $t+1$.
    \item Update $Q(s_t, a_t)$.
    \item Time $t \leftarrow t + 1$.
    \item Current state $s_t \leftarrow s_{t+1}$.
\end{enumerate}

The general workflow described above for Q-learning is similar to a "stop and wait" format: Until the reward is observed, and the state action pair's Q-value has been updated, RL learning agent cannot choose another action to perform. In our modified RL with aggregation of action selection, the workflow in each episode is as follows:

\begin{enumerate}
    \item Let current state be $s_0$, select an action $a_{0,1}$ which leads to a never reached state $s_1$, if the action is successfully performed.
    \item Add $a_{0,1}$ to a list $\{a_{0,1}, \dots\}$.
    \item Update the current state to be $s_1$ and repeat step 1 and 2. Stop repeating when the selected action leads to traffic demand’s destination, or to avoid stuck in a loop.
    \item Perform the actions $\{a_{0,1}, a_{1,2}, a_{2,3}, \dots\}$ in order.
    \item Observe and calculate rewards and store them in an ordered list.
    \item Updated corresponding state-action pairs, follow the order of performed actions.
\end{enumerate}

Compared with the unmodified workflow, in RLSR-Routing’s approach, the RL agent first selects all actions that will be performed during one episode. Then, the agent instructs the network to perform all the actions in order, and passively waits for returning QoS data which are used to calculate each action’s reward. The reason why we can aggregate action selection before updating Q-values are based on the following observations.

\textbf{Observation I:} Update $Q(s_t, a_t)$ does not affect action selection at state $s_{t+1}$, if selected action $a_{t+1}$ leads to a never-reached state during this episode.

\textbf{Observation II:} Update $Q(s_{t+1}, a_{t+1})$ does not affect updating $Q(s_t, a_t)$, if Q-values are updated in order of performed actions, and path is non-cyclic.

When an SARSA’s learning agent updates a state-action pair’s Q-value, it uses the following equation:

\begin{equation}
Q_{t+1}(s_t, a_t) = (1 - \alpha) \times Q_t(s_t, a_t) + \alpha \times (R + \gamma \times Q_t(s_{t+1}, a_{t+1}))
\end{equation}

$R$ stands for rewards, $\alpha$ represents the learning rate, and $\gamma$ indicates the importance of long-term rewards. Update of $(s_t, a_t)$'s Q-value at time $t+1$ depends on its old Q-value at time $t$ (i.e., $Q_t(s_t, a_t)$) and old Q-value for state-action pair at the next state (i.e., $Q_t(s_{t+1}, a_{t+1})$). When the path a.k.a. the list of actions, is non-cyclic, each state will only be included at most once in the path. If the update of Q-values follows the same order as actions performed, previously updated state-action pairs' Q-values will not be affected by updating the subsequent state-action pairs' Q-values.

Take topology $T_1$ as an example. Suppose now the aggregated selected actions are $\{a_{0,1}, a_{1,2}, a_{2,3}, a_{3,4}\}$, which is a non-cyclic path from node $0$ to node $4$. Starting from time $t$, the update of Q-values are as follows:


\begin{itemize}

\item $Q_{t+1}(s_0, a_{0,1}) = (1 - \alpha) \times Q_t(s_0, a_{0,1}) + \alpha \times (R_{0,1} + \gamma \times Q_t(s_1, a_{1,2})) $

\item $Q_{t+1}(s_1, a_{1,2}) = (1 - \alpha) \times Q_t(s_1, a_{1,2}) + \alpha \times (R_{1,2} + \gamma \times Q_t(s_2, a_{2,3})) $

\item $Q_{t+1}(s_2, a_{2,3}) = (1 - \alpha) \times Q_t(s_2, a_{2,3}) + \alpha \times (R_{2,3} + \gamma \times Q_t(s_3, a_{3,4})) $

\item $Q_{t+1}(s_3, a_{3,4}) = (1 - \alpha) \times Q_t(s_3, a_{3,4}) + \alpha \times (R_{3,4} + \gamma \times 1)$

\end{itemize}

Since action $a_{3,4}$ is the last one performed, here we just use $1$ to represent its next state-action pair's Q-value. As explained earlier, an update of any state-action pair's Q-value does not affect previously updated state-action pairs. Based on the above observations, we think that our modification will not affect the accuracy of Q-values updates. On the other hand, aggregate action selection can reduce network costs during the learning process, particularly communication costs between network nodes and the SDN controller which is explained below.

\begin{figure}[!htbp]
	\centering 
\includegraphics[width=0.4\textwidth]{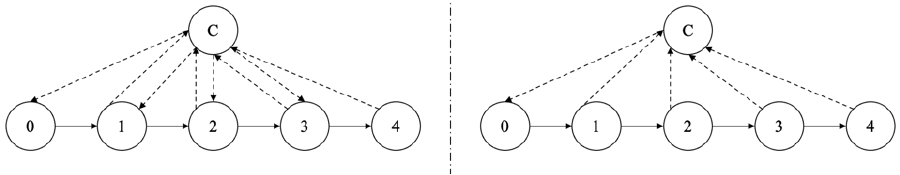}

    \caption{Comparing Number of Communications} 
        \label{fig:compare_com}
\end{figure}

Using the same example, Fig. \ref{fig:compare_com} illustrates the communication patterns between the controller (labelled as $C$) and network nodes under two different scenarios: without action selection aggregation (on the left) and with action selection aggregation (on the right). Without aggregation, the controller must send a new packet to the network for every newly selected action. In previous RL-based routing implementations like QAR in \cite{42}, the RL agent selects and performs only one action during each episode. For instance, even if the path from node $0$ to node $4$ is straightforward, when the current state is ``node $0$ holds the packet'', the RL agent can only instruct node $0$ to send the packet to node $1$. After this action is executed and the packet reaches node $1$ (updating the current state to ``node $1$ holds the packet''), the RL agent must select and perform a new action, sending an instruction to node $1$ regarding the next hop for the packet.

With aggregation and Source Routing (SR) technique, the controller can encode all selected actions into the packet's header, initiate routing, and then passively wait for Quality of Service (QoS) data to return from the network, assuming no packet loss occurs. Generally, if $n$ actions are to be performed during a single RL episode, the absence of action selection aggregation would result in $2n$ communications between the controller and nodes. With action selection aggregation, this number is reduced to $n+1$. Moreover, if the RL approach does not incorporate any techniques to prevent infinite loop formation, the number $n$ is more likely to equal the maximum Time-To-Live (TTL) value a packet is allowed within the network. Conversely, our modification avoids infinite loops by ensuring the RL agent is aware of the states already included in the path during the aggregated action selection process.

\subsubsection{RL Modification (Local and Global Reward):}

\begin{figure}[!htbp]
	\centering 
	\includegraphics[width=0.4\textwidth]{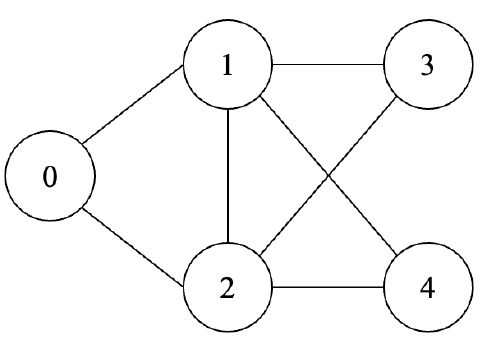}	
  \caption{A Simple Topology (T2)} 
	\label{fig:t2}
\end{figure}

In addition to the aggregate action selection process, we also aimed to use what the RL agent has learned during each episode more efficiently to speed up algorithm convergence. Taking QAR, proposed in \cite{42}, as an example, its RL algorithm workflow can be summarized as follows:

\begin{enumerate}
    \item Take source and destination node from a traffic demand.
    \item Initialize Q-table with all entries set to $0$.
    \item In each episode, repeat the Markov Decision Process.
    \item After RL converges, output the routing path from the updated Q-table.
\end{enumerate}

The Q-table is updated after the algorithm above converges, including but not limited to entries that together form a path with maximal rewards from the source to the destination. Since Q-values represent estimated cumulative rewards, the updated Q-table can be viewed as containing knowledge about some of the network's QoS status, assuming the reward function is based on those QoS parameters. However, as the RL agent handles a new traffic demand and reinitializes the Q-table’s entries to $0$, the previously learned knowledge about network QoS status is lost, requiring the RL agent to explore the network from scratch again.

Inspired by previous work, which provides some starting point during the RL agent’s initialization \cite{35}, we propose a “dual rewards scheme” for our modified RL algorithm to improve the efficacy of using RL results. Instead of one reward for one action, our algorithm calculates two rewards for each action: local and global rewards. The concept of local reward is similar to an action's reward in traditional Q-learning or SARSA: the RL agent uses local rewards to find actions that lead to maximum expected rewards (a.k.a. towards a preferred path defined by the user's customized QoS requirements). On the other hand, global rewards preserve the knowledge the RL agent learned about the network, which can be used in future learning processes.

As a result of using two kinds of reward, RLSR-Routing needs to construct two Q-tables: a local Q-table and a global Q-table. At the initialization phase, users of RLSR-Routing decide whether to use a global Q-table to initialize the local Q-table, instead of setting every entry to a random value like $0$. During the learning process, the agent only uses the local Q-table to select actions, since the main objective is to find a path for given traffic demands rather than learning overall network status. By adjusting weights of different QoS parameters, users of RLSR-Routing can customize the calculation of local rewards, and thus customize the evaluation of a path’s quality for different traffic demands. Meanwhile, the calculation of global rewards and updating the global Q-table are hidden from incoming traffic demands. To support potential future upgrades of RLSR-Routing that enables multi-threading, i.e., running multiple RL algorithm components to find a path for multiple traffic demands concurrently, we separated the global Q-table from the RL algorithm and made it another component of RLSR-Routing, as illustrated in Fig. \ref{fig:rsrl_component}.

Topology $T_2$ in Fig.\ref{fig:t2} is sufficient to demonstrate the advantage of using the dual rewards scheme. All links represent the two unidirectional links between two end nodes. The RL algorithm finds the best path between node $0$ and $3$ based on the number of hops and link reliability, and it has reached convergence. Assuming the final local and global Q-tables are as in Table \ref{tab:Local Q-table after Convergence (Hypothesized)} and \ref{tab:Global Q-table after Convergence (Hypothesized)} respectively.

\begin{table}[!htbp]
\centering
\caption{Local Q-table after Convergence (Hypothesized)}
\label{tab:Local Q-table after Convergence (Hypothesized)}
\begin{tabular}{c|ccccc}
\hline
       & Node 0 & Node 1 & Node 2 & Node 3 & Node 4 \\ \hline
Node 0 & X      & -1.5   & -1.8   & X      & X      \\
Node 1 & 0      & X      & -1.1   & -1.5   & -1.2   \\
Node 2 & 0      & -1.0   & X      & -0.8   & -1.3   \\
Node 3 & X      & 0      & 0      & X      & X      \\
Node 4 & X      & -1.2   & -1.1   & X      & X      \\ \hline
\end{tabular}
\end{table}

\begin{table}[!htbp]
\centering
\caption{Global Q-table after Convergence (Hypothesized)}
\label{tab:Global Q-table after Convergence (Hypothesized)}
\begin{tabular}{c|ccccc}
\hline
       & Node 0 & Node 1 & Node 2 & Node 3 & Node 4 \\ \hline
Node 0 & X      & -1.9   & -1.7   & X      & X      \\
Node 1 & 0      & X      & -1.5   & -1.4   & -1.1   \\
Node 2 & 0      & -0.2   & X      & -0.9   & -0.9   \\
Node 3 & X      & 0      & 0      & X      & X      \\
Node 4 & X      & -0.4   & -0.7   & X      & X      \\ \hline
\end{tabular}
\end{table}

Each row represents a state for Table \ref{tab:Local Q-table after Convergence (Hypothesized)} and \ref{tab:Global Q-table after Convergence (Hypothesized)}, and each column represents an action. The letter ``X'' indicates that the state-action pair does not exist, i.e., there is no direct link from the source to the destination of the node pair. A value of $0$ indicates that the agent did not explore the state-action pair during the learning process. Based on the local Q-table, a non-cyclic path with maximal reward from node $0$ to $3$ will be: $0 \rightarrow 1 \rightarrow 2 \rightarrow 3$. While the RL agent was exploring the best path for the traffic demand, it also updated the global Q-table based on link utilization, traffic intensity, and link reliability. Suppose now the RL agent needs to find a path for a traffic demand from node $0$ to node $4$, focusing on link utilization and traffic intensity instead of the number of hops. If the agent initializes all entries of the local Q-table to $0$, it has zero knowledge about network load-balancing status before the learning process begins. In contrast, if it uses the global Q-table to initialize the local Q-table, the agent will have some prior knowledge about the network, e.g., the first action should better choose to send the packet from node $0$ to $2$. We believe that in some cases, a priori knowledge of network QoS status will speed up algorithm convergence.

\subsection{RLSR-Routing Workflow}
In subsection \ref{subsec:rlsr-routing architecture} and \ref{subsec:state-action-reward}, we demonstrated RLSR-Routing’s architecture and explained the rationale behind some of our modifications to RL algorithm. In this subsection, we describe RLSR-Routing’s workflow in pseudo-code, especially on how our RL routing takes traffic demand from input and produces a user preferred path for a given demand. 

\begin{algorithm*}[!htbp]
\caption{Find Route using Reinforcement Learning}
\label{alg:RL_findRoute}
\begin{algorithmic}[1] 
\Function{RL\_findRoute}{$\text{TrafficDemand Demand} \newline \text{Boolean B, QoSWeights W, Hyperparameters H}$}
    \State \textit{// Initialization Phase}
    \State get Graph $G$ from RLSR-Routing; 
    \State get QTable $GT$ from RLSR-Routing; 
    \State get QoSWeights $Wd$ from RLSR-Routing; 
    \State get Hyperparameters $Hd$ from RLSR-Routing; 
    \State QTable $localT = \Call{InitLTable}{G, GT, B}$; 
    \State QoSWeights $wl = (W = \text{NULL}) ? Wd : W$; 
    \State Hyperparameters $hl = (H = \text{NULL}) ? Hd : H$; 
    \State $episodes = 0$
    \While{$episodes < hl.E$}
        \State Path $tempPath = \Call{FindTempPath}{Demand, localT, hl}$
        \State \Call{Push\_Path\_for\_Routing}{tempPath}
        \State QoSData[] $qData$, int $count = \Call{Collect\_QoS\_Data}{tempPath}$
        \State Reward[] $lRewards = \Call{CalculateLRewards}{qData, wl, count, Demand}$
        \State Reward[] $gRewards = \Call{CalculateGRewards}{qData, Wd, count}$
        \State \Call{UpdateTable}{localT, lRewards, hl, count}
        \State \Call{UpdateTable}{GT, gRewards, Hd, count}
        \State $episodes += 1$
    \EndWhile
    \State \Return \Call{FindFinalPath}{Demand, localT, hl}
\EndFunction
\end{algorithmic}
\end{algorithm*}

\textbf{RL\_findRoute()} in Algorithm \ref{alg:RL_findRoute} describes the overall workflow of the RL algorithm component: the user of RLSR-Routing initiates a routing request by providing traffic demand and optional customized factors. The RL algorithm explores the network and gradually learns the user-preferred path during

the learning phase. Once the algorithm converges or finishes the learning process, the final path is retrieved from the local Q-table. In practice, RLSR-Routing will be implemented and deployed on the SDN controller. By using the controller’s API, RLSR-Routing receives requests for finding paths for given traffic demands. Similarly, RLSR-Routing explores the network by using the controller’s API to send packets to SR-enabled network nodes. As network nodes route the packets based on their header-defined segments, RLSR-Routing waits for link-state information from network nodes. The final routing path for a traffic demand will be sent to the source node in the network, in order to achieve flow-based routing.

\subsection{RL Algorithm: Initialization}
Before RL agent starts learning process, it needs to extract information from user input as well as other RLSR-Routing components. This information is necessary to initialize local Q-table, QoS weights, hyperparameters and most importantly, the traffic demand to handle. 

\begin{algorithm*}[!htbp]
\caption{Initialize Local Q-Table}
\label{algo:Initialize Local Q-Table}
\begin{algorithmic}[1] 
\Function{InitLTable}{$\text{Graph } G, \text{QTable } GT, \text{Boolean } B$}
    \If{$B$ \textbf{is} $\text{true}$}
        \State $\text{QTable localT} = \Call{DeepCopy}{GT};$
        \State \Return $\text{localT};$
    \EndIf
    \State $\text{QTable localT} = \text{new QTable}(G.\Call{numberOfNodes}{})$
    \State $\text{Link}[][] \text{links} = G.\Call{getLinkMatrix}{}$
    \For{$i = 0$ \textbf{to} $\text{links.length} - 1$}
        \For{$j = 0$ \textbf{to} $\text{links}[0].\text{length} - 1$}
            \If{$\text{link}[i][j] == \text{NULL}$}
                \State $\text{localT}[i][j] = -0x80000000;$
            \Else
                \State $\text{localT}[i][j] = 0;$
            \EndIf
        \EndFor
    \EndFor
    \State \Return $\text{localT};$
\EndFunction
\end{algorithmic}
\end{algorithm*}

\textbf{InitLTable()} in Algorithm \ref{algo:Initialize Local Q-Table} is a helper function for initializing the local Q-table. Both global and local Q-tables are implemented as a 2-dimensional array of doubles, where the array entry $[i][j]$ represents the Q-value for the state-action pair ``at node $i$, sends packet to node $j$''. If the user opts to apply the global Q-table, the helper function creates a deep copy of the global Q-table as the starting point for the local Q-table. It is recognized that, in some cases, users of RLSR-Routing may wish to apply the global Q-table but not copy every entry from the global to the local Q-table exactly. Future enhancements may include more options for applying the global Q-table for initialization. 

On the other hand, users can initialize the local Q-table from scratch: for every existing link, the corresponding value is initialized to 0. For links that do not exist, the corresponding state-action pair's value is initialized to a minimal integer value. During initialization steps, the RL agent also determines whether to use default QoS weights and/or hyperparameters, based on user input. The structure of QoS weights and hyperparameters is presented in Table \ref{tab:Structure of QoS weights} and \ref{tab:Structure of Hyperparameters} respectively.

\begin{table}[!htbp]
\centering
\caption{Structure of QoS weights}
\label{tab:Structure of QoS weights}
\begin{tabular}{|p{3cm}|p{5cm}|}
\hline
\textbf{Variable} & \textbf{Description} \\ 
\hline
\texttt{double lConstant;} & Constant used when calculating global reward \\ 
\hline
\texttt{double gConstant;} & Constant used when calculating local reward \\ \hline
\texttt{double Wc;} & Hop-count reward’s weight \\ \hline
\texttt{double Wt;} & Transmission rate reward’s weight \\ \hline
\texttt{double Wr;} & Link reliability reward’s weight \\ \hline
\texttt{double Wi;} & Traffic intensity reward’s weight \\ \hline
\texttt{double Wu;} & Link utilization reward’s weight \\
\hline
\end{tabular}
\end{table}

\begin{table}[!htbp]
\centering
\caption{Structure of Hyperparameters}
\label{tab:Structure of Hyperparameters}
\begin{tabular}{|p{3cm}|p{5cm}|}
\hline
\textbf{Variable} & \textbf{Description} \\ 
\hline
\texttt{double $\varepsilon$;} & For $\varepsilon$-greedy action selection \\ \hline
\texttt{double $\alpha$;} & Learning rate \\ \hline
\texttt{double $\gamma$;} & Importance of long-term rewards \\ \hline
\texttt{int TTL;} & Maximum number of hops \\ \hline
\texttt{int E;} & Number of training episodes \\ 
\hline
\end{tabular}
\end{table}

Users can provide customized weights to emphasize the importance of different factors, and variables $gConstant$ and $lConstant$ depend on these weights, as illustrated below. Hyperparameters contain fields for RL’s action selection and Q-value updating ($\varepsilon$, $\alpha$, $\gamma$) and for adjusting training duration (TTL, $E$). 

\begin{equation}
lConstant = Wc + Wt + Wr + Wi + Wu + 0.1
\end{equation}

\begin{equation}
gConstant = Wr + Wi + Wu
\end{equation}

\subsection{RL Algorithm: Actions Selection}

\begin{figure}[!htbp]
	\centering 
	\includegraphics[width=0.4\textwidth]{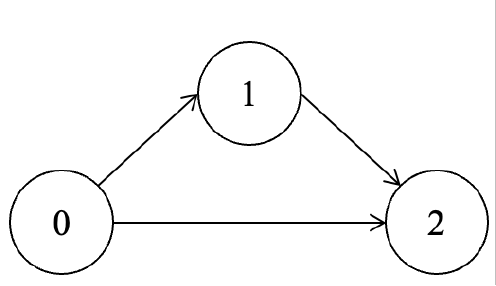}
   \caption{Edge Case Topology (T3)} 
	\label{fig:t3}
\end{figure}

\begin{algorithm*}[!htbp]
\caption{Find Temporary Path}
\label{alg:FindTempPath}
\begin{algorithmic}[1]
\Function{FindTempPath}{$\text{TrafficDemand } D, \text{QTable } L, \text{Hyperparameters } H$}
    \State $\text{Link}[][] \text{links} = \text{get link matrix from Graph } G, \text{from RLSR-Routing};$
    \State $\text{int}[] \text{visitedNodes} = \text{new int}[G.\text{numberOfNodes}];$
    \State initialize all $\text{visitedNodes}$ entries to $0$;
    \State $\text{int currentId} = D.\text{srcId};$
    \State $\text{int nextId} = -1;$
    \State $\text{Path tempPath};$
    \State $\text{int remainTTL} = H.\text{TTL};$
    \While{$\text{remainTTL} > 0 \textbf{ and } \text{nextId} \neq D.\text{dstId}$}
        \State $\text{Node}[] \text{neighbors} = \text{get current node's neighbors from Graph};$
        \State $\text{Node}[] \text{unvisited} = \text{neighbors} \cap (\text{visitedNodes with entry} = 0);$
        \If{$\text{unvisited is empty}$}
            \State break; 
        \EndIf
        \If{$\text{Probability } P < H.\varepsilon$}
            \State nextId = randomly selects one unvisited neighbor; 
        \Else
            \State nextId = one with highest Q-value from $L$; 
        \EndIf
        \State add selected action(next hop) to $\text{tempPath}$, and updates related fields;
        \State $\text{unvisited[nextId]} = 1;$
        \State $\text{currentId} = \text{nextId};$ 
        \State $\text{remainTTL} = \text{remainTTL} - 1;$
    \EndWhile
    \State \Return $\text{tempPath};$
\EndFunction
\end{algorithmic}
\end{algorithm*}

\textbf{FindTempPath()} in Algorithm \ref{alg:FindTempPath} describes how the RL algorithm uses aggregate action selection to produce a non-cyclic path. The function keeps track of every node that has been added to the temporary path, and because RLSR-Routing assigns each network node an integer ID that starts from 0, \texttt{visitedNodes[i]} is enough to express whether node $i$ has been included. Starting from the traffic demand's source node,\texttt{FindTempPath()} builds a consecutive path by selecting from the current node's unvisited neighbors. The path building process ends due to one of the following: 1) a dead-end, i.e., all current node's neighbors have been included in \texttt{tempPath}. 2) the path length reaches the time-to-live a packet is allowed to travel in the network. 3) the path has reached the traffic demand's destination node.

It should be noted that \texttt{FindTempPath()} follows an action selection policy throughout its running process, even if the traffic demand's destination node is one of the current node's neighbors. Considering topology T3 in Fig. \ref{fig:t3}, all links are unidirectional, and now there is a traffic demand from node 0 to node 2, with QoS consideration on link utilization. Suppose link $l_{0,2}$ currently has a 99\% utilization rate, whereas link $l_{0,1}$ and $l_{1,2}$ has 0\% utilization. Although node 2 is adjacent to node 0, the preferred path is $0 \rightarrow 1 \rightarrow 2$ instead of $0 \rightarrow 2$.

\subsection{RL Algorithm: Perform Actions \& Observe QoS Data}

\texttt{Push\_Path\_for\_Routing()} and \texttt{Collect\_QoS\_Data()} in Algorithm \ref{alg:RL_findRoute} represent the process of generating a packet, encoding \texttt{tempPath} into the packet’s header, sending the packet to the network via the SDN controller for routing, and waiting for related link-state QoS data collected by the SDN controller. In this study, we only focused on simulating the overall workflow. However, during actual deployment on SDN architecture, these functions will be implemented by other components of RLSR-Routing. Modifications to the SR protocol may also be necessary so that packets generated by RLSR-Routing in the learning process are distinguished from other packets in the network. Therefore, network devices will only send related QoS data to the SDN controller when they process packets created by RLSR-Routing’s RL algorithm. Another approach to modifying the protocol could involve adding a field called ``SEND QoS DATA'' on all packet headers. Network switches would then only send QoS data to the controller for certain packets (such as those created by the RL algorithm during the learning process) that have the ``SEND QoS DATA'' field set.

Although we assumed that communications between network devices and the controller are reliable, the returned QoS data may be out-of-order due to factors like physical distance between a node and the controller. \texttt{Collect\_QoS\_Data()} will sort incoming data in order of \texttt{tempPath}’s added links. Moreover, \texttt{Collect\_QoS\_Data()} should be able to detect packet loss through mechanisms such as a timer. When packet loss occurs, it signifies that one performed action failed to complete (i.e., transform the current state to the next state), and no QoS data for the failed action will be received. The action performing process will terminate, and this failed action becomes the last action performed. Subsequently, \texttt{Collect\_QoS\_Data()} should actively send a request to the controller to explicitly require related link-state information. To distinguish a failed action from an action without packet loss, a field in the \texttt{QoSData} structure, \texttt{hasLost}, has been added to indicate whether packet loss occurred when the action was performed.

\subsection{RL Algorithm: Rewards Calculation}
The local and global reward calculations are given by:
\begin{align}
    R_{\text{local}} &= W_c \times R_{\text{hop}} + W_t \times R_{\text{transmission}} + W_r \times R_{\text{reliability}} \nonumber \\
    &\quad + W_i \times R_{\text{inten-est}} + W_u \times R_{\text{util-est}} - l_{\text{Constant}} \\
    R_{\text{global}} &= W_r \times R_{\text{reliability}} + W_i \times R_{\text{inten}} + W_u \times R_{\text{util}} - g_{\text{Constant}} 
\end{align}

Eq. 11 and 12 are calculations of local reward and global reward, respectively. Each individual QoS factor's reward is calculated based on Eq. 1 – 7, and two constant values calculations ($l_{\text{Constant}}$ and $g_{\text{Constant}}$) are based on Eq. 9 and 10. In our traffic routing problem, local rewards can be viewed as a reflection of path quality for a given traffic demand. On the other hand, global rewards are used to learn the current network status. Thus, the current traffic intensity (Rinten) and current link utilization (Rutil) are used to calculate global rewards. Eq. 11 ensures all local rewards are no greater than -0.1, and Eq. 12 ensures all global rewards are no greater than 0. 

\begin{algorithm*}[!htbp]
\caption{Calculate Local Rewards}
\begin{algorithmic}[1]
\Function{CalculateLRewards}{QoSData $D$, QoSWeights $W$, int $Count$, TrafficDemand $Demand$}
    \State Reward[] $lRewards$; 
    \For{$i = 0$; $i < Count - 1$; $i++$}
        \State $lRewards[i].srcId = D.srcId$;
        \State $lRewards[i].dstId = D.dstId$;
        \State $lRewards[i].actionSuccess = \text{True}$;
        \State $lRewards[i].value = \text{use Eq. 11, } D, W \text{ to calculate}$;
    \EndFor
    \If{$D.hasLost$ \textbf{OR} $D.dstId \neq Demand.dstId$}
        \State $lRewards[Count-1].actionSuccess = \text{False}$;
        \State $lRewards[Count-1].value = -W.lConstant$;
    \Else
        \State $lRewards[Count-1].actionSuccess = \text{True}$;
        \State $lRewards[Count-1].value = \text{use Eq. 11, } D, W \text{ to calculate}$;
    \EndIf
    \State \Return $lRewards$;
\EndFunction
\end{algorithmic}
\end{algorithm*}

\begin{algorithm*}[!htbp]
\caption{Calculate Global Rewards}
\begin{algorithmic}[1]
\Function{CalculateGRewards}{QoSData $D$, QoSWeights $W$, int $Count$}
    \State Reward[] $gRewards$; 
    \For{$i = 0$; $i < Count - 1$; $i++$}
        \State $gRewards[i].srcId = D.srcId$;
        \State $gRewards[i].dstId = D.dstId$;
        \State $gRewards[i].actionSuccess = \text{True}$;
        \State $gRewards[i].value = \text{use Eq. 12, } D, W \text{ to calculate}$;
    \EndFor
    \If{$D.hasLost$}
        \State $gRewards[Count-1].actionSuccess = \text{False}$;
        \State $gRewards[Count-1].value = -W.gConstant$;
    \Else
        \State $gRewards[Count-1].actionSuccess = \text{True}$;
        \State $gRewards[Count-1].value = \text{use 4.12, } D, W \text{ to calculate}$;
    \EndIf
    \State \Return $gRewards$;
\EndFunction
\end{algorithmic}
\end{algorithm*}

\begin{algorithm*}[!htbp]
\caption{UpdateTable function}
\begin{algorithmic}[1]
\Function{UpdateTable}{$QTable\ T,\ Rewards\ R,\ Hyperparameters\ H, \ int\ Count$}
    \State //update Q-values for the first till the second last performed actions
    \For{$int\ i = 0;\ i < Count - 1;\ i++$}
        \State $int\ state = R[i].srcId;$
        \State $int\ action = R[i].dstId;$
        \State $T[state][action] =$ use Eq. 8 to update Q-value;
    \EndFor
    \State $int\ state = R[Count-1].srcId;$
    \State $int\ action = R[Count-1].dstId;$
    \State $double\ nextQVal = 0;$
    \If{$R[count-1].actionSuccess == False$}
        \State $T[state][action] += R[Count-1].value;$
    \Else
        \State $T[state][action] =$ use Eq. 8 and nextQVal to update;
    \EndIf
\EndFunction
\end{algorithmic}
\end{algorithm*}

\textit{CalculateLRewards()} and \textit{CalculateGRewards()} show the pseudocode for calculating local and global rewards. Both functions have a similar workflow, applying Eq. 11 and 12 to calculate local/global rewards for the first performed action until the second last. For the last performed action, packet loss may occur, necessitating a different approach to calculate the rewards and potentially applying a punishment for unsuccessfully performed actions, so that RL agent learns to avoid select those actions in the future. However, definition of “unsuccessfully performed last action” is different from the local or global reward’s perspective. For local rewards calculation, the last action is considered unsuccessful if it failed to deliver the packet to its traffic demand’s destination, even if no packet loss happened. For global rewards calculation, the last action is considered unsuccessful only when packet loss occurred. In addition to assigning punishment values for unsuccessful actions, RL algorithm adopted in RLSR-Routing has different ways to update related Q-values.

\subsection{RL Algorithm: Tables Update}
The RL algorithm uses the \texttt{UpdateTable()} function for both local and global Q-table updates. For performed actions that are guaranteed to be successful (from the first action to the second last), the function applies SARSA’s typical function (Eq. 8) to update corresponding Q-values. For the last action, classified as unsuccessfully performed, the associated Q-value accumulates the penalty by adding the reward’s value. The reason \texttt{UpdateTable()} does not use Eq. 8 for the unsuccessful actions' Q-value update is discussed subsequently.

\begin{figure}[!htbp]
	\centering 
	\includegraphics[width=0.4\textwidth]{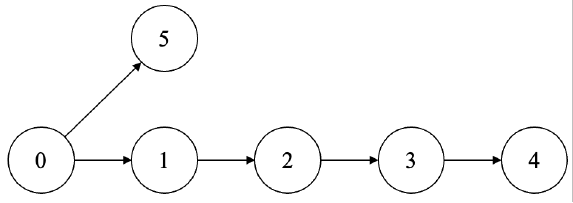}
   \caption{Edge Case Topology (T4)} 
	\label{fig:t4}
\end{figure}

Topology T4 in Fig. \ref{fig:t4} illustrates an edge case in which using Eq. 8 to update an unsuccessful action's Q-value may cause a problem in RLSR-Routing. Consider a scenario where a traffic demand is to send packets from node 0 to node 4. Every successfully performed action has a local reward value of $-2$; whereas, unsuccessfully performed ones have a local reward value of $-3$. The RL agent uses $\alpha = \gamma = 1$, $\varepsilon = 0.5$, and $TTL = 16$ for hyperparameters. No packet loss occurred during the learning process. After running enough episodes that the local Q-table converges, the final local Q-table should look like this:

Based on the setting of the RL algorithm, when node 0 chooses to send the packet to node 5, such action is regarded as unsuccessful by local rewards calculation. This is because from node 5, the RL agent cannot find a never visited next-hop node to deliver the packet to the traffic demand’s destination. If we use Eq. 8 to update $Q(s_0, a_{0,5})$, the value will converge at $-3$ based on our settings above. As a result, at node 0, choosing action $a_{0,5}$ has a higher cumulative reward than choosing action $a_{0,1}$ ($-3$ vs $-8$), even if sending a packet to node 1 is the only way to reach the final destination node 4. In contrast, the accumulative penalty mechanism in \texttt{UpdateTable()} ensures that the more times an action is performed unsuccessfully, the lower Q-value it will receive. For example, consider the updates to the Q-values as the agent repeatedly chooses action \(a_{0,5}\):
\begin{itemize}
    \item The first time the agent chooses \(a_{0,5}\), \(Q(s_0, a_{0,5})\) will become \(-3\).
    \item The second time \(Q(s_0, a_{0,5})\) will become \(-6\).
    \item The third time \(Q(s_0, a_{0,5})\) will become \(-9\),
    \item and so on.
\end{itemize}

\subsection{RL Algorithm: Return final result}

After all learning episodes are completed, the RL algorithm uses \texttt{FindFinalPath()} to retrieve the final routing path from the source node to the destination node of the input traffic demand. \texttt{FindFinalPath()} executes \texttt{FindTempPath()}, but with Hyperparameters' \(\varepsilon\) value set to \(0\) – i.e., using a greedy selection policy. With an adequate number of learning episodes, the final path should be non-cyclic and successfully reaches the destination node, with the highest estimated cumulative reward from local Q-table. However, the final path may not be the best solution based on the user’s QoS requirements, and one possible reason could be the action selection policy used during the learning process.

%% file: experiments.tex
We provided two ways to create a network topology for performing experiments. First is to use our implemented methods to write code, which step by step creates a graph representation of a network, adds nodes and links, and possibly assigns the initial used bandwidth of certain links. Another way is to write a JSON file with a specified format, which can be parsed using an open-source JSON parser. 

RLSR-Routing is a framework that directly addresses path finding issues in SDN, without requiring prior knowledge of the network (e.g., link weights) or external libraries to train a neural network. Deployment of RLSR-Routing should be flexible as long as it can interact with the SDN controller and parse JSON file. This study mainly focused on the RL algorithm component’s implementation, with additional code to support network simulation. All code is written in JAVA but transfer to other programming languages should not be a complicated task. This section presents our comparative study and validation experiments’ results with discussion.

\subsection{Settings for Hyperparameters}
Authors who proposed QAR in \cite{42} conducted experiments to study the effects of learning rate and the importance of long-term rewards. We followed their guidelines and unless specified explicitly, the hyperparameters used in this experiments are summarized in Table \ref{table:hyperparameters_values}.

\begin{table}[!htbp]
\centering
\caption{Hyperparameters, their values, and description}
\label{table:hyperparameters_values}
\begin{tabular}{|p{1.5cm}|p{1cm}|p{5cm}|}
\hline
\textbf{Parameter} & \textbf{Value} & \textbf{Description}                                                                                       \\ \hline
$\epsilon$         & 0              & Exploration rate in the $\epsilon$-greedy strategy, set to 0 for no exploration.                        \\ \hline
$\alpha$           & 0.9            & Learning rate, controls how quickly the algorithm updates its knowledge with new information.          \\ \hline
$\gamma$           & 0.9            & Discount factor, determines the importance of future rewards in the reinforcement learning algorithm.  \\ \hline
$\text{TTL}$       & 32             & Time to Live, may represent the number of steps or iterations before termination in specific contexts. \\ \hline
$E$                & 75             & Represent episodes.                    \\ \hline
\end{tabular}

\end{table}

\begin{figure}[!htbp]
	\centering 
	\includegraphics[width=9cm]{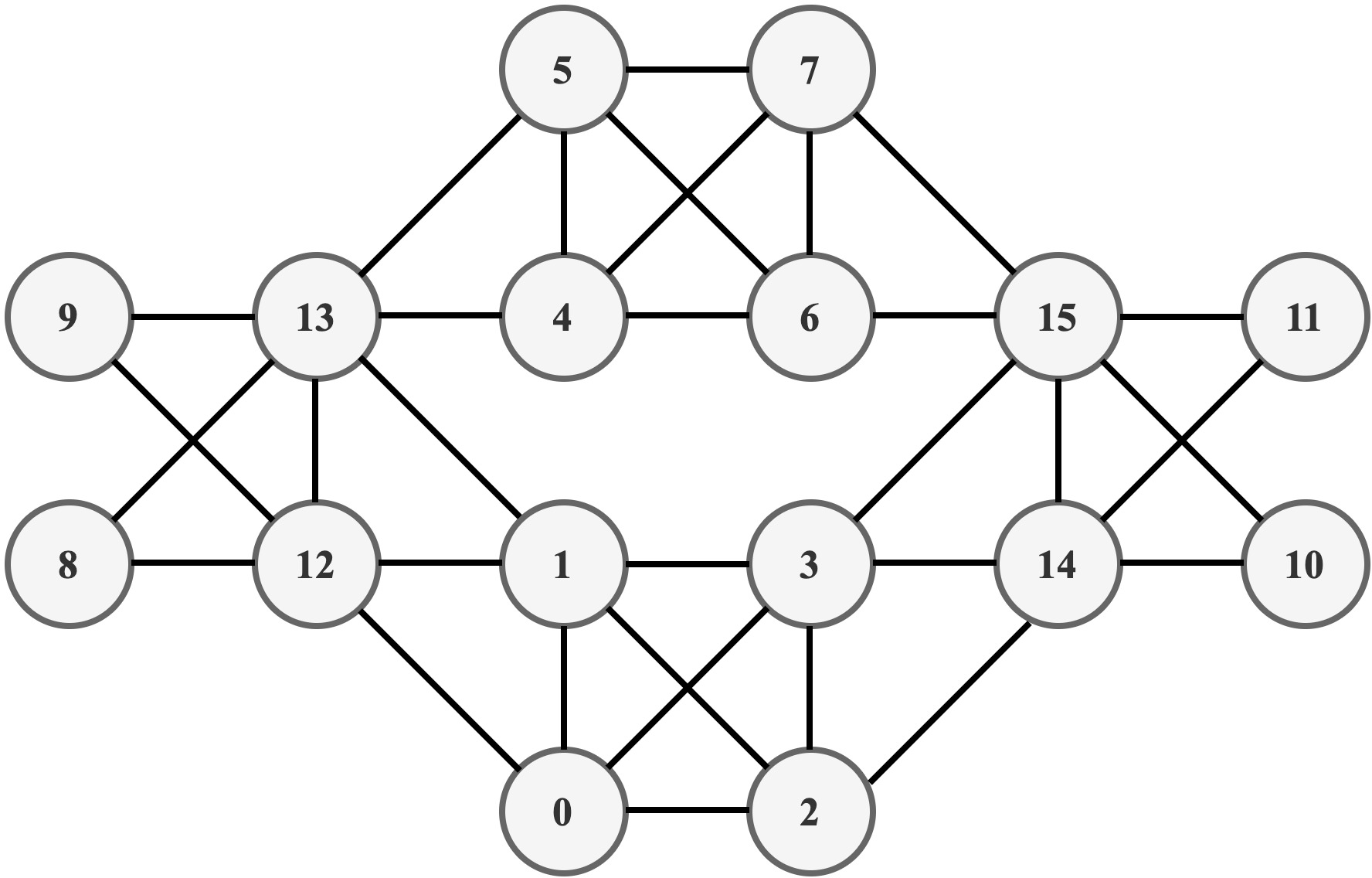}	
   \caption{16 Nodes Topology (T7))} 
	\label{fig:t7}
\end{figure}

\subsection{Comparative Study with Non-RL routing}

We performed a comparative study between \textbf{RLSR-Routing} and a non RL-based routing algorithm, \textbf{NR-Routing}, which is used by a telecom solutions provider. Similar to our proposed RLSR-Routing, NR-Routing is able to assign paths for given traffic flows between source and destination pairs; however, it uses a non-RL based agent to explore the traffic engineering space. The agent places traffic demands greedily (from largest to smallest) and/or randomly on the network capacity using multi-path segment routing based traffic engineering. Due to confidentiality issues, we cannot provide additional information about NR-Routing’s implementation or other details. The overall objective is to compare the two algorithms' load balancing ability in terms of minimizing maximum link utilization over a given network. The testing network topology \textit{T7} is shown in Fig. \ref{fig:t7}, with 16 nodes and each pair of adjacent nodes has two unidirectional links connecting them. All links have a 2.0Mb/s maximum bandwidth, 100\% reliability rate, and all nodes have a 20Mb/s processing rate.

NR-Routing is an executable program embodying two main functions:
\begin{itemize}
    \item Generation of traffic engineering problems consisting of network topologies and sets of traffic demands to place on the network capacity.
    \item Optimization of traffic demand engineering by placing demands on paths in the network with segment routing based traffic steering techniques.
\end{itemize}

We first randomly generated networks and demands and let NR-Routing assign one or more paths for each demand. These generated paths were saved as a JSON file. We then removed unwanted demands, adjusted the amount of data for the remaining, and then used them as NR-Routing’s input for load balancing. This time, the final output is saved in another JSON file to be used for comparison. The traffic and paths assignments for selected tunnels presented in Table \ref{table:traffic_distribution} that we used in NR-Routing program’s load balancing test.

\begin{table}[!htbp]
\centering
\caption{Traffic distribution among nodes}
\label{table:traffic_distribution}
\begin{tabular}{|c|c|c|c|}
\hline
\textbf{Source} & \textbf{Destination} & \textbf{Total Traffic} & \textbf{Number of Paths} \\ \hline
4                  & 14               & 400kb/s                & 2                        \\ \hline
1                  & 5                & 400kb/s                & 2                        \\ \hline
0                  & 6                & 1.0Mb/s                & 5                        \\ \hline
13                 & 11               & 600kb/s                & 4                        \\ \hline
3                  & 13               & 600kb/s                & 3                        \\ \hline
3                  & 4                & 1.2Mb/s                & 6                        \\ \hline
\end{tabular}

\end{table}

\begin{table}[!htbp]
\centering
\caption{RLSR-Routing Generated Paths for Traffic Demands}
\label{table:RLSR-Routing Generated Paths for Traffic Demands}
\begin{tabular}{cl}
\toprule
\textbf{Source-Destination} & \textbf{Path} \\
\midrule
4 to 14 & i. 4 $\rightarrow$ 7 $\rightarrow$ 15 $\rightarrow$ 14 \\
        & ii. 4 $\rightarrow$ 6 $\rightarrow$ 15 $\rightarrow$ 14 \\
\midrule
1 to 5  & i. 1 $\rightarrow$ 13 $\rightarrow$ 5 \\
        & ii. 1 $\rightarrow$ 13 $\rightarrow$ 5 \\
\midrule
0 to 6  & i. 0 $\rightarrow$ 3 $\rightarrow$ 15 $\rightarrow$ 6 \\
        & ii. 0 $\rightarrow$ 2 $\rightarrow$ 14 $\rightarrow$ 15 $\rightarrow$ 6 \\
        & iii. 0 $\rightarrow$ 12 $\rightarrow$ 13 $\rightarrow$ 4 $\rightarrow$ 6 \\
        & iv. 0 $\rightarrow$ 3 $\rightarrow$ 15 $\rightarrow$ 7 $\rightarrow$ 6 \\
        & v. 0 $\rightarrow$ 1 $\rightarrow$ 3 $\rightarrow$ 15 $\rightarrow$ 6 \\
\midrule
13 to 11 & i. 13 $\rightarrow$ 1 $\rightarrow$ 2 $\rightarrow$ 14 $\rightarrow$ 11 \\
         & ii. 13 $\rightarrow$ 1 $\rightarrow$ 3 $\rightarrow$ 14 $\rightarrow$ 11 \\
         & iii. 13 $\rightarrow$ 4 $\rightarrow$ 7 $\rightarrow$ 15 $\rightarrow$ 11 \\
         & iv. 13 $\rightarrow$ 5 $\rightarrow$ 6 $\rightarrow$ 15 $\rightarrow$ 11 \\
\midrule
3 to 13 & i. 3 $\rightarrow$ 1 $\rightarrow$ 13 \\
        & ii. 3 $\rightarrow$ 0 $\rightarrow$ 12 $\rightarrow$ 13 \\
        & iii. 3 $\rightarrow$ 1 $\rightarrow$ 13 \\
\midrule
3 to 4  & i. 3 $\rightarrow$ 15 $\rightarrow$ 7 $\rightarrow$ 4 \\
        & ii. 3 $\rightarrow$ 2 $\rightarrow$ 1 $\rightarrow$ 12 $\rightarrow$ 8 $\rightarrow$ 13 $\rightarrow$ 4 \\
        & iii. 3 $\rightarrow$ 14 $\rightarrow$ 15 $\rightarrow$ 7 $\rightarrow$ 4 \\
        & iv. 3 $\rightarrow$ 15 $\rightarrow$ 6 $\rightarrow$ 4 \\
        & v. 3 $\rightarrow$ 1 $\rightarrow$ 13 $\rightarrow$ 4 \\
        & vi. 3 $\rightarrow$ 0 $\rightarrow$ 12 $\rightarrow$ 13 $\rightarrow$ 4 \\
\bottomrule
\end{tabular}
\end{table}

\begin{table}[ht]
\centering
\caption{NR-Routing’s Links with Traffic}
\label{tab:NR-Routing’s Links with Traffic}
\begin{tabular}{|c|c|c|}
\hline
\textbf{No.} & \textbf{Path} & \textbf{Bandwidth (BD)} \\
\hline
1 & R3(14) $\rightarrow$ R4(15) & 8000000.0 \\
2 & R4(15) $\rightarrow$ R3(14) & 4000000.0 \\
3 & B2(5) $\rightarrow$ B3(6) & 316666.0 \\
4 & R1(12) $\rightarrow$ R2(13) & 14333330.0 \\
5 & B3(6) $\rightarrow$ R4(15) & 966666.0 \\
6 & R4(15) $\rightarrow$ B3(6) & 500000.0 \\
7 & A2(1) $\rightarrow$ R2(13) & 500000.0 \\
8 & A1(0) $\rightarrow$ R1(12) & 933333.0 \\
9 & A4(3) $\rightarrow$ R4(15) & 466666.0 \\
10 & R2(13) $\rightarrow$ B2(5) & 866666.0 \\
11 & B1(4) $\rightarrow$ B3(6) & 516666.0 \\
12 & B3(6) $\rightarrow$ B1(4) & 300000.0 \\
13 & A4(3) $\rightarrow$ R3(14) & 466666.0 \\
14 & R2(13) $\rightarrow$ B1(4) & 1066666.0 \\
15 & A2(1) $\rightarrow$ R1(12) & 500000.0 \\
16 & A3(2) $\rightarrow$ R3(14) & 3333330.0 \\
17 & B1(4) $\rightarrow$ B4(7) & 3500000.0 \\
18 & B4(7) $\rightarrow$ B1(4) & 3000000.0 \\
19 & A4(3) $\rightarrow$ A1(0) & 6000000.0 \\
20 & A1(0) $\rightarrow$ A3(2) & 3333330.0 \\
21 & A4(3) $\rightarrow$ A2(1) & 6000000.0 \\
22 & B4(7) $\rightarrow$ R4(15) & 5000000.0 \\
23 & R4(15) $\rightarrow$ B4(7) & 3000000.0 \\
24 & B2(5) $\rightarrow$ B4(7) & 1500000.0 \\
25 & R4(15) $\rightarrow$ PE4(11) & 6000000.0 \\
\hline
\end{tabular}

\end{table}

\begin{table}[ht]
\centering
\caption{RLSR-Routing's Links with Traffic}
\label{tab:RLSR-Routing's Links with Traffic}
\begin{tabular}{|c|c|c|}
\hline
 No. & Path & Bandwidth (BD) \\
\hline
 1 & 0$\rightarrow$1 & 200000.0 \\
 2 & 0$\rightarrow$2 & 200000.0 \\
 3 & 0$\rightarrow$3 & 400000.0 \\
 4 & 0$\rightarrow$12 & 600000.0 \\
 5 & 1$\rightarrow$2 & 150000.0 \\
 6 & 1$\rightarrow$3 & 350000.0 \\
 7 & 1$\rightarrow$12 & 200000.0 \\
 8 & 1$\rightarrow$13 & 1000000.0 \\
 9 & 2$\rightarrow$1 & 200000.0 \\
 10 & 2$\rightarrow$14 & 350000.0 \\
 11 & 3$\rightarrow$0 & 400000.0 \\
 12 & 3$\rightarrow$1 & 600000.0 \\
 13 & 3$\rightarrow$2 & 200000.0 \\
 14 & 3$\rightarrow$14 & 350000.0 \\
 15 & 3$\rightarrow$15 & 1600000.0 \\
 16 & 4$\rightarrow$6 & 400000.0 \\
 17 & 4$\rightarrow$7 & 350000.0 \\
 18 & 5$\rightarrow$6 & 150000.0 \\
 19 & 6$\rightarrow$4 & 200000.0 \\
 20 & 6$\rightarrow$15 & 350000.0 \\
 21 & 7$\rightarrow$4 & 400000.0 \\
 22 & 7$\rightarrow$6 & 200000.0 \\
 23 & 7$\rightarrow$15 & 350000.0 \\
 24 & 8$\rightarrow$13 & 200000.0 \\
 25 & 12$\rightarrow$8 & 200000.0 \\
 26 & 12$\rightarrow$13 & 600000.0 \\
 27 & 13$\rightarrow$1 & 300000.0 \\
 28 & 13$\rightarrow$4 & 950000.0 \\
 29 & 13$\rightarrow$5 & 550000.0 \\
 30 & 14$\rightarrow$11 & 300000.0 \\
 31 & 14$\rightarrow$15 & 400000.0 \\
 32 & 15$\rightarrow$6 & 800000.0 \\
 33 & 15$\rightarrow$7 & 600000.0 \\
 34 & 15$\rightarrow$11 & 300000.0 \\
 35 & 15$\rightarrow$14 & 400000.0 \\
\hline
\end{tabular}
\end{table}

\begin{figure}[ht]
	\centering 

	\includegraphics[width=9cm]{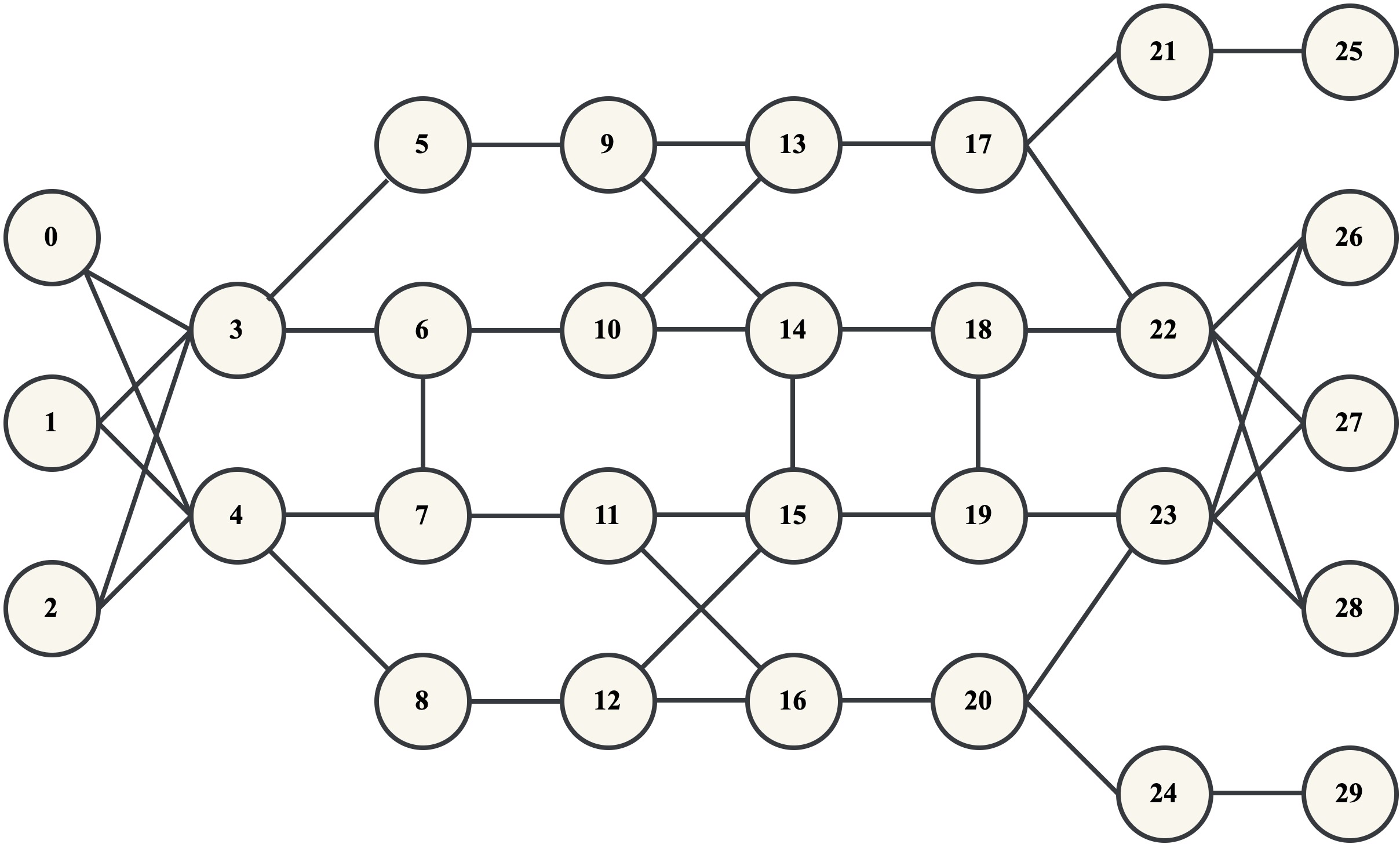}	
     \caption{30 Nodes Topology (T8)} 
	\label{fig:t8}
\end{figure}

For RLSR-Routing, it interpreted each tunnel as $n$ traffic demands, where $n$ equals the number of paths assigned to the tunnel by NR-Routing. Every traffic demand from the same tunnel split the tunnel’s total traffic equally. For example, for tunnels from node 3 to 4, each of the six traffic demands had 200kb/s traffic. The RL algorithm only considered link utilization when calculating local rewards during the learning process. After the RL algorithm found a path for one traffic demand, the demand’s traffic was placed over the network along the path. These steps were repeated until the RL algorithm found paths for all traffic demands and placed all demand traffic over the network.

\subsection{Routing Paths by RLSR-Routing}
The paths generated for each tunnel (source and destination pair) by RLSR-Routing are meticulously listed in Table \ref{table:RLSR-Routing Generated Paths for Traffic Demands}, showcasing the algorithm's ability to produce valid, loop-free routing paths that consist of consecutive nodes connecting the specified source and destination. All paths produced by RLSR-Routing are validated as efficient routing paths; they are loop-free and comprise a consecutive series of nodes linking the source and destination nodes for specified traffic demands.

\subsection{Links Utilizations}
Table \ref{tab:NR-Routing’s Links with Traffic} and \ref{tab:RLSR-Routing's Links with Traffic} show the links that placed traffic on it by NR-Routing and RLSR-Routing, respectively. For NR-Routing, traffic assigned on different links of the network was saved in the JSON output file. We extracted the information and saved in another text file. Although the same network topology was used in both NR-Routing and RLSR-Routing, the labels for each node used in NR-Routing are different. Therefore, we included the translated node’s ID in brackets, for example, node with ID R3 in NR-Routing is node with ID (14) in RLSR-Routing. For NR-Routing, the maximum traffic assigned to one link is 1.433333Mb/s, thus the maximum link utilization in NR-Routing’s output is  1.433333 / 2.0 = 71.67\%. For RLSR-Routing, the maximum traffic assigned to one link is 1.0Mb/s, thus the maximum link utilization in RLSR-Routing’s links assignment is 1.0 / 2.0 = 50.0\%.

\begin{table*}[ht]
\centering
\caption{Final Path for Traffic Demands}
\label{tab:final_path_traffic_demands}
\begin{tabular}{|c|l|}
\hline
\textbf{Traffic Demand} & \multicolumn{1}{c|}{\textbf{Final Path}} \\ \hline
0 $\rightarrow$ 26 & 0 $\rightarrow$ 4 $\rightarrow$ 7 $\rightarrow$ 6 $\rightarrow$ 10 $\rightarrow$ 14 $\rightarrow$ 18 $\rightarrow$ 19 $\rightarrow$ 23 $\rightarrow$ 26 \\ \hline
1 $\rightarrow$ 26 & 1 $\rightarrow$ 4 $\rightarrow$ 7 $\rightarrow$ 6 $\rightarrow$ 10 $\rightarrow$ 14 $\rightarrow$ 18 $\rightarrow$ 19 $\rightarrow$ 23 $\rightarrow$ 26 \\ \hline
2 $\rightarrow$ 26 & 2 $\rightarrow$ 4 $\rightarrow$ 7 $\rightarrow$ 6 $\rightarrow$ 10 $\rightarrow$ 14 $\rightarrow$ 18 $\rightarrow$ 19 $\rightarrow$ 23 $\rightarrow$ 26 \\ \hline
0 $\rightarrow$ 27 & 0 $\rightarrow$ 4 $\rightarrow$ 7 $\rightarrow$ 6 $\rightarrow$ 10 $\rightarrow$ 14 $\rightarrow$ 18 $\rightarrow$ 19 $\rightarrow$ 23 $\rightarrow$ 27 \\ \hline
1 $\rightarrow$ 27 & 1 $\rightarrow$ 4 $\rightarrow$ 7 $\rightarrow$ 6 $\rightarrow$ 10 $\rightarrow$ 14 $\rightarrow$ 18 $\rightarrow$ 19 $\rightarrow$ 23 $\rightarrow$ 27 \\ \hline
2 $\rightarrow$ 27 & 2 $\rightarrow$ 4 $\rightarrow$ 7 $\rightarrow$ 6 $\rightarrow$ 10 $\rightarrow$ 14 $\rightarrow$ 18 $\rightarrow$ 19 $\rightarrow$ 23 $\rightarrow$ 27 \\ \hline
0 $\rightarrow$ 28 & 0 $\rightarrow$ 4 $\rightarrow$ 7 $\rightarrow$ 6 $\rightarrow$ 10 $\rightarrow$ 14 $\rightarrow$ 18 $\rightarrow$ 19 $\rightarrow$ 23 $\rightarrow$ 28 \\ \hline
1 $\rightarrow$ 28 & 1 $\rightarrow$ 4 $\rightarrow$ 7 $\rightarrow$ 6 $\rightarrow$ 10 $\rightarrow$ 14 $\rightarrow$ 18 $\rightarrow$ 19 $\rightarrow$ 23 $\rightarrow$ 28 \\ \hline
2 $\rightarrow$ 28 & 2 $\rightarrow$ 4 $\rightarrow$ 7 $\rightarrow$ 6 $\rightarrow$ 10 $\rightarrow$ 14 $\rightarrow$ 18 $\rightarrow$ 19 $\rightarrow$ 23 $\rightarrow$ 28 \\ \hline
\end{tabular}
\end{table*}

\begin{figure*}[ht]
\centering
\subfloat[TempPath length for Demand $0 \rightarrow 26$]{\includegraphics[width=0.3\textwidth]{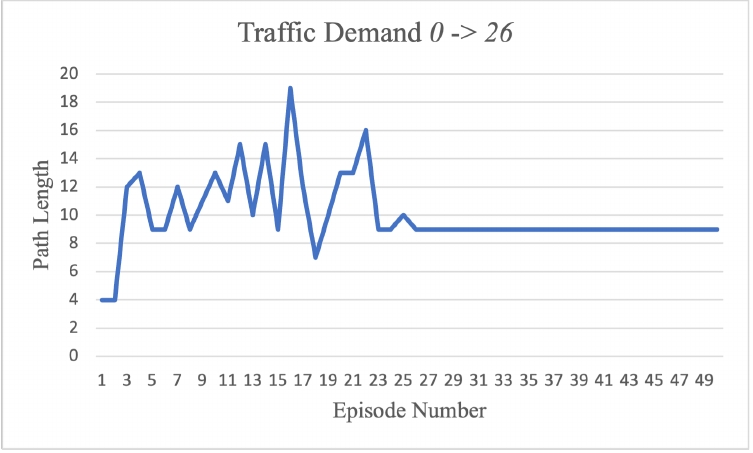}}\quad
\subfloat[TempPath length for Demand $1 \rightarrow 26$]{\includegraphics[width=0.3\textwidth]{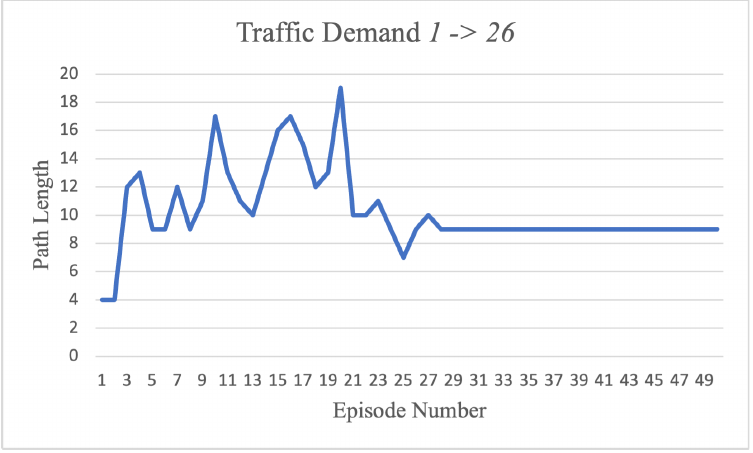}}\quad
\subfloat[TempPath length for Demand $2 \rightarrow 26$]{\includegraphics[width=0.3\textwidth]{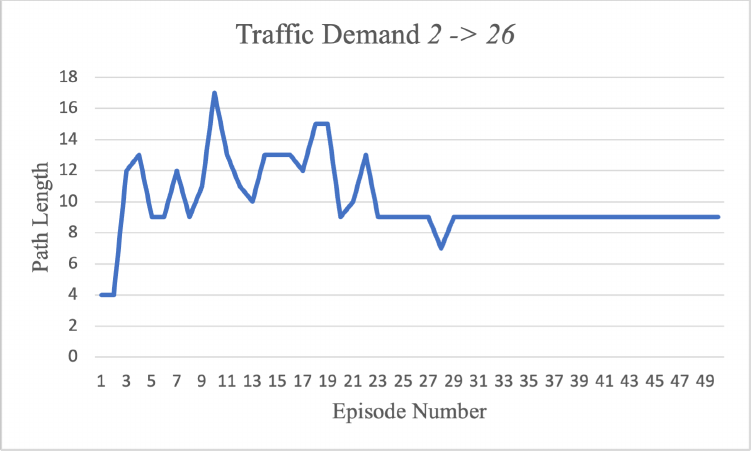}}\\
\subfloat[TempPath length for Demand $0 \rightarrow 27$]{\includegraphics[width=0.3\textwidth]{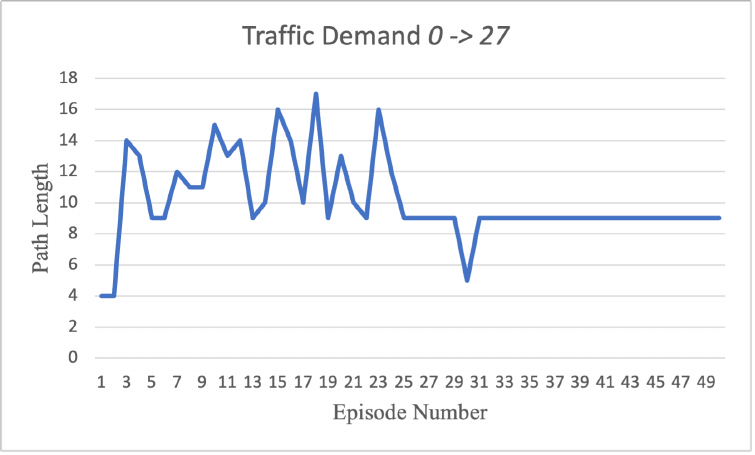}}\quad
\subfloat[TempPath length for Demand $0 \rightarrow 27$]{\includegraphics[width=0.3\textwidth]{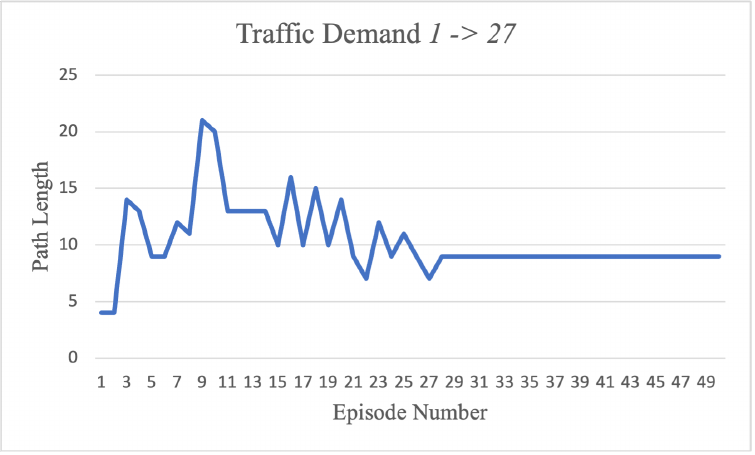}}\quad
\subfloat[TempPath length for Demand $0 \rightarrow 27$]{\includegraphics[width=0.3\textwidth]{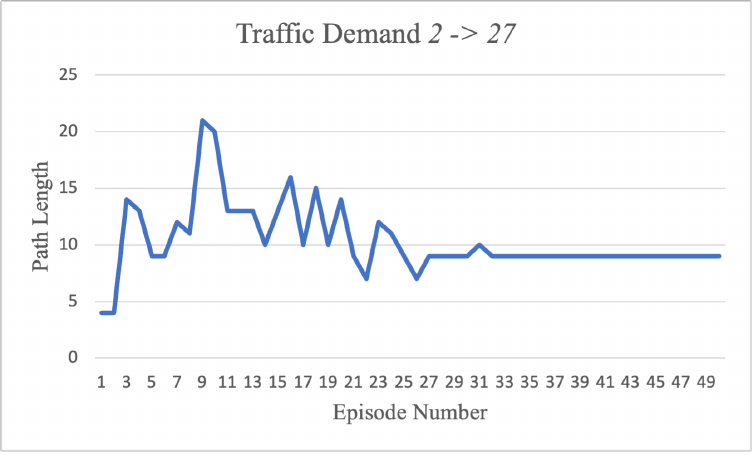}}\\
\subfloat[TempPath length for Demand $0 \rightarrow 28$]{\includegraphics[width=0.3\textwidth]{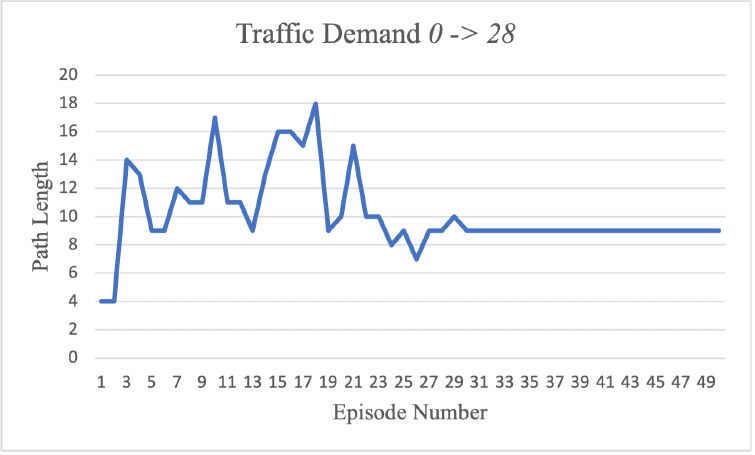}}\quad
\subfloat[TempPath length for Demand $0 \rightarrow 28$]{\includegraphics[width=0.3\textwidth]{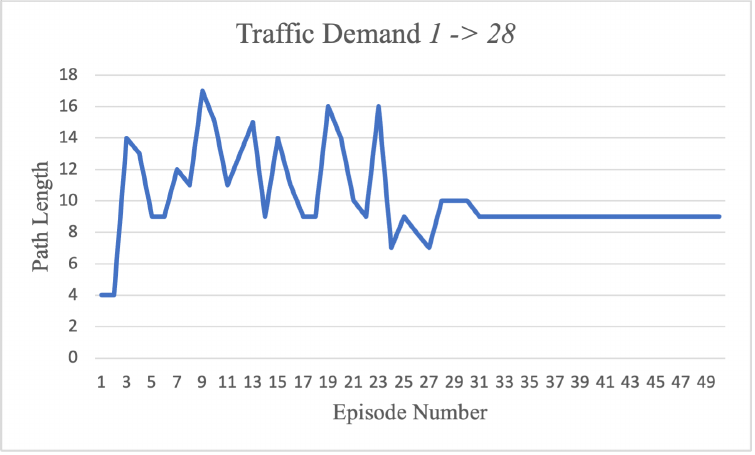}}\quad
\subfloat[TempPath length for Demand $0 \rightarrow 28$]{\includegraphics[width=0.3\textwidth]{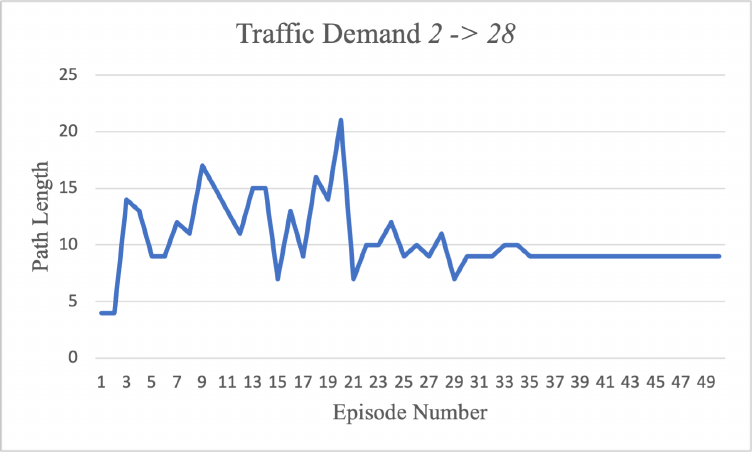}}
\caption{Evolution of TempPath Lengths During Early Learning Episodes: Illustrates the length of temporary paths (TempPath) for various demands across the first 50 episodes in RLSR-Routing's learning process. Despite initial challenges in reaching destinations within the first 20 episodes, the agent rapidly adapted, ensuring paths without loops and mostly under 20 nodes. Convergence to efficient routing paths was typically achieved by the 30th episode, highlighting RLSR-Routing's effectiveness in learning optimal paths for traffic demands.}
\label{fig:pathvsepisode}
\end{figure*}

\begin{table*}[ht]
\centering
\caption{Final Paths by Test Group I ($\gamma = 0.3$)}
\label{table:Test Group I }
\begin{tabular}{|c|c|}
\hline
Traffic Demand & Final Path \\
\hline
0 $\rightarrow$ 26 & 0 $\rightarrow$ 4 $\rightarrow$ 7 $\rightarrow$ 6 $\rightarrow$ 10 $\rightarrow$ 14 $\rightarrow$ 18 $\rightarrow$ 19 $\rightarrow$ 23 $\rightarrow$ 26 \\
1 $\rightarrow$ 26 & 1 $\rightarrow$ 4 $\rightarrow$ 7 $\rightarrow$ 6 $\rightarrow$ 10 $\rightarrow$ 14 $\rightarrow$ 18 $\rightarrow$ 19 $\rightarrow$ 23 $\rightarrow$ 26 \\
2 $\rightarrow$ 26 & 2 $\rightarrow$ 4 $\rightarrow$ 7 $\rightarrow$ 6 $\rightarrow$ 10 $\rightarrow$ 14 $\rightarrow$ 18 $\rightarrow$ 19 $\rightarrow$ 23 $\rightarrow$ 26 \\
0 $\rightarrow$ 27 & 0 $\rightarrow$ 4 $\rightarrow$ 7 $\rightarrow$ 6 $\rightarrow$ 10 $\rightarrow$ 14 $\rightarrow$ 18 $\rightarrow$ 19 $\rightarrow$ 23 $\rightarrow$ 27 \\
1 $\rightarrow$ 27 & 1 $\rightarrow$ 4 $\rightarrow$ 7 $\rightarrow$ 6 $\rightarrow$ 10 $\rightarrow$ 14 $\rightarrow$ 18 $\rightarrow$ 19 $\rightarrow$ 23 $\rightarrow$ 27 \\
2 $\rightarrow$ 27 & 2 $\rightarrow$ 4 $\rightarrow$ 7 $\rightarrow$ 6 $\rightarrow$ 10 $\rightarrow$ 14 $\rightarrow$ 18 $\rightarrow$ 19 $\rightarrow$ 23 $\rightarrow$ 27 \\
0 $\rightarrow$ 28 & 0 $\rightarrow$ 4 $\rightarrow$ 7 $\rightarrow$ 6 $\rightarrow$ 10 $\rightarrow$ 14 $\rightarrow$ 18 $\rightarrow$ 19 $\rightarrow$ 23 $\rightarrow$ 28 \\
1 $\rightarrow$ 28 & 1 $\rightarrow$ 4 $\rightarrow$ 7 $\rightarrow$ 6 $\rightarrow$ 10 $\rightarrow$ 14 $\rightarrow$ 18 $\rightarrow$ 19 $\rightarrow$ 23 $\rightarrow$ 28 \\
2 $\rightarrow$ 28 & 2 $\rightarrow$ 4 $\rightarrow$ 7 $\rightarrow$ 6 $\rightarrow$ 10 $\rightarrow$ 14 $\rightarrow$ 18 $\rightarrow$ 19 $\rightarrow$ 23 $\rightarrow$ 28 \\
\hline
\end{tabular}
\end{table*}

\begin{table*}[ht]
  \centering
    \caption{Final Paths by Test Group II ($\gamma = 0.5$)}
  \label{tab:Test Group II }
  \begin{tabular}{|c|c|}
    \hline
    Traffic Demand & Final Path \\
    \hline
    0 $\rightarrow$ 26 & 0 $\rightarrow$ 4 $\rightarrow$ 7 $\rightarrow$ 6 $\rightarrow$ 10 $\rightarrow$ 14 $\rightarrow$ 18 $\rightarrow$ 19 $\rightarrow$ 23 $\rightarrow$ 26 \\
    \hline
    1 $\rightarrow$ 26 & 1 $\rightarrow$ 4 $\rightarrow$ 7 $\rightarrow$ 6 $\rightarrow$ 10 $\rightarrow$ 14 $\rightarrow$ 18 $\rightarrow$ 19 $\rightarrow$ 23 $\rightarrow$ 26 \\
    \hline
    2 $\rightarrow$ 26 & 2 $\rightarrow$ 4 $\rightarrow$ 7 $\rightarrow$ 6 $\rightarrow$ 10 $\rightarrow$ 14 $\rightarrow$ 18 $\rightarrow$ 19 $\rightarrow$ 23 $\rightarrow$ 26 \\
    \hline
    0 $\rightarrow$ 27 & 0 $\rightarrow$ 4 $\rightarrow$ 7 $\rightarrow$ 6 $\rightarrow$ 10 $\rightarrow$ 14 $\rightarrow$ 18 $\rightarrow$ 19 $\rightarrow$ 23 $\rightarrow$ 27 \\
    \hline
    1 $\rightarrow$ 27 & 1 $\rightarrow$ 4 $\rightarrow$ 7 $\rightarrow$ 6 $\rightarrow$ 10 $\rightarrow$ 14 $\rightarrow$ 18 $\rightarrow$ 19 $\rightarrow$ 23 $\rightarrow$ 27 \\
    \hline
    2 $\rightarrow$ 27 & 2 $\rightarrow$ 4 $\rightarrow$ 7 $\rightarrow$ 6 $\rightarrow$ 10 $\rightarrow$ 14 $\rightarrow$ 18 $\rightarrow$ 19 $\rightarrow$ 23 $\rightarrow$ 27 \\
    \hline
    0 $\rightarrow$ 28 & 0 $\rightarrow$ 4 $\rightarrow$ 7 $\rightarrow$ 6 $\rightarrow$ 10 $\rightarrow$ 14 $\rightarrow$ 18 $\rightarrow$ 19 $\rightarrow$ 23 $\rightarrow$ 28 \\
    \hline
    1 $\rightarrow$ 28 & 1 $\rightarrow$ 4 $\rightarrow$ 7 $\rightarrow$ 6 $\rightarrow$ 10 $\rightarrow$ 14 $\rightarrow$ 18 $\rightarrow$ 19 $\rightarrow$ 23 $\rightarrow$ 28 \\
    \hline
    2 $\rightarrow$ 28 & 2 $\rightarrow$ 4 $\rightarrow$ 7 $\rightarrow$ 6 $\rightarrow$ 10 $\rightarrow$ 14 $\rightarrow$ 18 $\rightarrow$ 19 $\rightarrow$ 23 $\rightarrow$ 28 \\
    \hline
  \end{tabular}

\end{table*}

\begin{table*}[ht]
\centering
\caption{Final Paths by Test Group III ($\gamma = 0.7$)}
\label{tab:Test Group III }
\begin{tabular}{|c|c|}
\hline
Traffic Demand& Final Path \\
\hline
0 $\rightarrow$ 26 & 0 $\rightarrow$ 4 $\rightarrow$ 7 $\rightarrow$ 6 $\rightarrow$ 10 $\rightarrow$ 14 $\rightarrow$ 18 $\rightarrow$ 19 $\rightarrow$ 23 $\rightarrow$ 26 \\
\hline
1 $\rightarrow$ 26 & 1 $\rightarrow$ 4 $\rightarrow$ 7 $\rightarrow$ 6 $\rightarrow$ 10 $\rightarrow$ 14 $\rightarrow$ 18 $\rightarrow$ 19 $\rightarrow$ 23 $\rightarrow$ 26 \\
\hline
2 $\rightarrow$ 26 & 2 $\rightarrow$ 4 $\rightarrow$ 7 $\rightarrow$ 6 $\rightarrow$ 10 $\rightarrow$ 14 $\rightarrow$ 18 $\rightarrow$ 19 $\rightarrow$ 23 $\rightarrow$ 26 \\
\hline
0 $\rightarrow$ 27 & 0 $\rightarrow$ 4 $\rightarrow$ 7 $\rightarrow$ 6 $\rightarrow$ 10 $\rightarrow$ 14 $\rightarrow$ 18 $\rightarrow$ 19 $\rightarrow$ 23 $\rightarrow$ 27 \\
\hline
1 $\rightarrow$ 27 & 1 $\rightarrow$ 4 $\rightarrow$ 7 $\rightarrow$ 6 $\rightarrow$ 10 $\rightarrow$ 14 $\rightarrow$ 18 $\rightarrow$ 19 $\rightarrow$ 23 $\rightarrow$ 27 \\
\hline
2 $\rightarrow$ 27 & 2 $\rightarrow$ 4 $\rightarrow$ 7 $\rightarrow$ 6 $\rightarrow$ 10 $\rightarrow$ 14 $\rightarrow$ 18 $\rightarrow$ 19 $\rightarrow$ 23 $\rightarrow$ 27 \\
\hline
0 $\rightarrow$ 28 & 0 $\rightarrow$ 4 $\rightarrow$ 7 $\rightarrow$ 6 $\rightarrow$ 10 $\rightarrow$ 14 $\rightarrow$ 18 $\rightarrow$ 19 $\rightarrow$ 23 $\rightarrow$ 28 \\
\hline
1 $\rightarrow$ 28 & 1 $\rightarrow$ 4 $\rightarrow$ 7 $\rightarrow$ 6 $\rightarrow$ 10 $\rightarrow$ 14 $\rightarrow$ 18 $\rightarrow$ 19 $\rightarrow$ 23 $\rightarrow$ 28 \\
\hline
2 $\rightarrow$ 28 & 2 $\rightarrow$ 4 $\rightarrow$ 7 $\rightarrow$ 6 $\rightarrow$ 10 $\rightarrow$ 14 $\rightarrow$ 18 $\rightarrow$ 19 $\rightarrow$ 23 $\rightarrow$ 28 \\
\bottomrule
\end{tabular}

\end{table*}

\begin{table*}[ht]
\centering
\caption{Final Paths by Test Group IV ($\gamma = 0.9$)}
\label{tab:Test Group IV }
\begin{tabular}{|c|c|}
\hline
Traffic Demand & Final Path \\
\hline
0 $\rightarrow$ 26 & 0 $\rightarrow$ 4 $\rightarrow$ 7 $\rightarrow$ 6 $\rightarrow$ 10 $\rightarrow$ 14 $\rightarrow$ 18 $\rightarrow$ 19 $\rightarrow$ 23 $\rightarrow$ 26 \\ \hline
1 $\rightarrow$ 26 & 1 $\rightarrow$ 4 $\rightarrow$ 7 $\rightarrow$ 6 $\rightarrow$ 10 $\rightarrow$ 14 $\rightarrow$ 18 $\rightarrow$ 19 $\rightarrow$ 23 $\rightarrow$ 26 \\ \hline
2 $\rightarrow$ 26 & 2 $\rightarrow$ 4 $\rightarrow$ 7 $\rightarrow$ 6 $\rightarrow$ 10 $\rightarrow$ 14 $\rightarrow$ 18 $\rightarrow$ 19 $\rightarrow$ 23 $\rightarrow$ 26 \\ \hline
0 $\rightarrow$ 27 & 0 $\rightarrow$ 4 $\rightarrow$ 7 $\rightarrow$ 6 $\rightarrow$ 10 $\rightarrow$ 14 $\rightarrow$ 18 $\rightarrow$ 19 $\rightarrow$ 23 $\rightarrow$ 27 \\ \hline
1 $\rightarrow$ 27 & 1 $\rightarrow$ 4 $\rightarrow$ 7 $\rightarrow$ 6 $\rightarrow$ 10 $\rightarrow$ 14 $\rightarrow$ 18 $\rightarrow$ 19 $\rightarrow$ 23 $\rightarrow$ 27 \\ \hline
2 $\rightarrow$ 27 & 2 $\rightarrow$ 4 $\rightarrow$ 7 $\rightarrow$ 6 $\rightarrow$ 10 $\rightarrow$ 14 $\rightarrow$ 18 $\rightarrow$ 19 $\rightarrow$ 23 $\rightarrow$ 27 \\ \hline
0 $\rightarrow$ 28 & 0 $\rightarrow$ 4 $\rightarrow$ 7 $\rightarrow$ 6 $\rightarrow$ 10 $\rightarrow$ 14 $\rightarrow$ 18 $\rightarrow$ 19 $\rightarrow$ 23 $\rightarrow$ 28 \\ \hline
1 $\rightarrow$ 28 & 1 $\rightarrow$ 4 $\rightarrow$ 7 $\rightarrow$ 6 $\rightarrow$ 10 $\rightarrow$ 14 $\rightarrow$ 18 $\rightarrow$ 19 $\rightarrow$ 23 $\rightarrow$ 28 \\ \hline
2 $\rightarrow$ 28 & 2 $\rightarrow$ 4 $\rightarrow$ 7 $\rightarrow$ 6 $\rightarrow$ 10 $\rightarrow$ 14 $\rightarrow$ 18 $\rightarrow$ 19 $\rightarrow$ 23 $\rightarrow$ 28 \\
\hline
\end{tabular}
\end{table*}

\begin{table*}[!htbp]
\centering
\caption{Episode Number when RL Converged}
\label{tab:Episode Number when RL Converged}
\begin{tabular}{|c|c|c|c|c|c|}
\hline
\textbf{Demand} & \textbf{Control} & \textbf{Test I $\gamma = 0.3$} & \textbf{Test II $\gamma = 0.5$} & \textbf{Test III $\gamma = 0.7$} & \textbf{Test IV $\gamma = 0.9$} \\ \hline
0 $\rightarrow$ 26 &25 & 25&25 & 25& 25\\ \hline
1 $\rightarrow$ 26 & 27&26 &25 & 21&18 \\ \hline
2 $\rightarrow$ 26 &30 &28 &38 & 17&17 \\ \hline
0 $\rightarrow$ 27 &30 & 21& 24&23 &20 \\ \hline
1 $\rightarrow$ 27 &27& 24&25 &21 &20 \\ \hline
2 $\rightarrow$ 27 & 31& 25& 26& 24&25 \\ \hline
0 $\rightarrow$ 28 &29 &22 &25 &27 &21 \\ \hline
1 $\rightarrow$ 28 &30 & 26& 28&29 &22 \\ \hline
2 $\rightarrow$ 28 &34 &34 &32 &32 &24 \\ \hline
\end{tabular}
\end{table*}

\begin{figure}[!htbp]
\centering
\includegraphics[width=8cm]{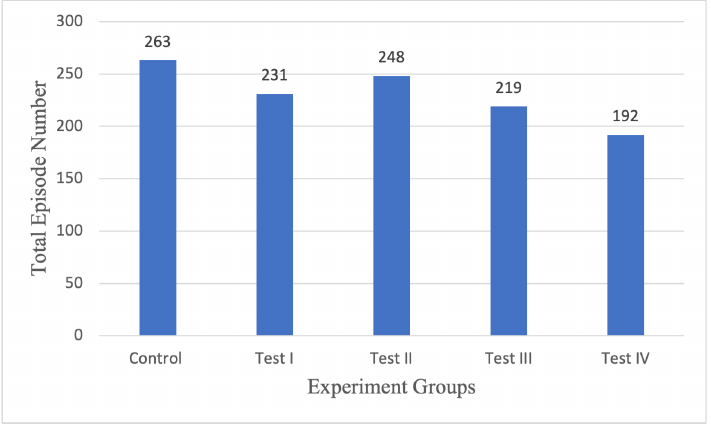}
\caption{Total Episode Numbers when RL Converged for All Demands}
\label{fig:total_episodes}
\end{figure}

\subsection{Cost of Exploration During Learning Process}
\label{subsec:cost of Exploration During Learning Process}
We used a 30 nodes network as shown in Fig. ~\ref{fig:t8} to test the efficacy of RLSR-Routing in the learning process. In topology T8, each pair of adjacent nodes are connected by two unidirectional links to support bi-directional communications. All nodes have a 100Mb/s processing rate, and all links have a 10Mb/s maximum bandwidth. Before any traffic demand's data been placed over the network, links used bandwidth are as follows:
\begin{itemize}
    \item Link \(l_{6,7}\), \(l_{7,6}\), \(l_{18,19}\), \(l_{19,18}\): 0Mb/s used bandwidth;
    \item Link \(l_{1,3}\), \(l_{2,3}\): 5Mb/s used bandwidth;
    \item Link \(l_{0,3}\), \(l_{4,8}\), \(l_{7,11}\), \(l_{10,13}\), \(l_{18,22}\): 9Mb/s used bandwidth;
    \item The rest of the links: 1Mb/s used bandwidth.
\end{itemize}
In this experiment, we gave RLSR-Routing nine traffic demands, each with 0.1Mb/s estimated traffic. The RL algorithm found one path for every given traffic demand, with traffic intensity and link utilization as considered QoS parameters (\(W_i = W_u = 1\)). A global Q-table was not used during this experiment since the experiment was not focused on convergence speed. After one path was returned from the RL algorithm, corresponding traffic demand data was placed over the network (i.e., all links on the path added 0.1Mb/s to their used bandwidth).

\subsubsection{Final Paths}
Table~\ref{tab:final_path_traffic_demands} summarizes the final routing path returned by RLSR-Routing for every traffic demand. All routing paths are valid: every node in a path is only included once, and every path reaches the specified destination. During the learning process, RLSR-Routing gradually explored the network, so that highly utilized links (like link \(l_{0,3}\), \(l_{4,8}\), \(l_{7,11}\), \(l_{10,13}\), \(l_{18,22}\)) and nodes with relatively high traffic intensity (like node 3, 8, 11, 13, 22) are excluded in final paths. As a result, all the paths include the same set of nodes, node 4, 7, 6, 10, 14, 18, 19, 23 as intermediate nodes between source and destination. Although these are not the shortest paths in terms of hop-count, they are the user preferred paths in terms of link utilization and traffic intensity.

\subsubsection{Path Length vs Episode Number}
Fig. \ref{fig:pathvsepisode} demonstrated when RLSR-Routing was finding the path for each traffic demand, the length (i.e., number of hops) of temporary paths (denote as TempPath) generated in the first 50 episodes. None of the temporary paths would result in a packet inside a loop, and most of the paths have less than 20 nodes included. Although at the beginning of the learning process, especially during the first 20 episodes, some temporary paths failed to reach the desired destination, the agent quickly learned to avoid choosing actions that lead to a dead end. For most of the testing traffic demands, RLSR-Routing was able to converge around 30 episodes; after that, the temporary paths produced in each episode are the same as the final paths.

\subsubsection{Global Q-table and Effects of \(\gamma\) Value}
We present our study about whether using global Q-table speeds up algorithm convergence. We applied the same network topology (T8) and the same initial links used bandwidth as in subsection \ref{subsec:cost of Exploration During Learning Process}. The same set of traffic demands were given to RLSR-Routing and the user preferred path was defined by the same QoS weights. The only difference is that RLSR-Routing used a global Q-table to initialize local Q-table. To study the effect of \(\gamma\) (importance of long term reward) on applying global Q-table, we set up five experiment groups. The control group from subsection \ref{subsec:cost of Exploration During Learning Process} did not use global Q-table; test groups I to IV applied \(\gamma\) value 0.3, 0.5, 0.7, and 0.9 to update global Q-table’s entries. The final paths for each traffic demand and the number of episodes RLSR-Routing used to converge were recorded for further analysis.
Table \ref{table:Test Group I } to \ref{tab:Test Group IV } illustrate different test groups’ final paths for given traffic demands.

\subsubsection{RL Convergence Speed}
Table~\ref{tab:Episode Number when RL Converged} shows for each traffic demand, the number of episodes different experiment groups used to reach convergence. Compared with the control group, in most cases, test groups used fewer episodes to reach convergence for the same traffic demand. Fig. ~\ref{fig:total_episodes} summarizes the total episode numbers for each group to reach convergence on 9 traffic demands. All test groups used fewer total episode numbers to find paths for all traffic demands than the control group. Test group IV, which used \(\gamma\) value 0.9, has the fastest convergence speed: it took 192 episodes to find paths for all given traffic demands.

%% file: conclusion.tex
We developed a RL-based routing algorithm, RLSR-Routing, which works on SR-enabled SDN. We modified SARSA, an on-policy RL algorithm, by aggregating action selection and setting a dual rewards scheme. To handle potential packet loss or failure to reach the desired destination during the learning process, we added additional steps than the normal function to update Q-tables for failed actions. Experiment results indicate that RLSR-Routing can find user preferred paths for given traffic demand, based on customized QoS considerations.

Our research contribution is listed as follows:

\begin{itemize}
    \item RLSR-Routing directly applies RL in the path finding process, without a prior knowledge of the network. It does not rely on additional inputs such as pre-defined link weights; or a set of pre-calculated paths between a source and destination pair to find the user preferred path of a traffic demand.
    \item We further reduced the number of communications required between SDN controller and network data planes by exploiting Segment Routing, compared with previous RL-based routing in SDN.
    \item We ensured that no packet would be routed in a loop during the RLSR-Routing learning process. 
    \item We gave users the ability to customize QoS weights to define what is their preferred path.
    \item Our proposed framework can reuse previously learned knowledge to speed up algorithm convergence.
\end{itemize}

Action selection based on a greedy approach may not find the best path in some cases. Relative medium size networks and small number of traffic demands tested may be the reasons that we only observed user preferred final paths in experiment results. Another issue is that RLSR-Routing does not support finding paths for multiple paths in parallel. When working on a set of traffic demands, RLSR-Routing only focuses on finding the best path for one traffic demand at a time, without considering the effect of placing traffic on the best path on other traffic demands that have not been assigned a path. To provide better demonstrations of paths assignments and links utilizations by RLSR-Routing and NR-Routing, we only used a subset of randomly generated demands during the comparative study. In practice, RLSR-Routing should assign paths for all traffic demands from the users of the network.

As explained earlier, several upgrades can further improve RLSR-Routing’s performance. For example, a better action selection policy, and more flexibility for users to decide how to apply a global Q-table to initiate a local Q-table. For the comparative study, we plan to test our RLSR-Routing in a more practical setting of traffic engineering problem, such as a 1000 nodes networks with thousands of traffic demands. In addition, future experiment should test RLSR-Routing’s ability to find user preferred path in terms of other QoS parameters, such as link reliability, instead of just using link utilization and traffic intensity. Finally, we will modify the whole RL algorithm component so that it can compute the user preferred path for a traffic demand matrix in parallel.